\documentclass[a4paper,11pt]{article}

\usepackage{bbold,amssymb,amsmath}

\newcommand{\ns}{\normalsize}

\numberwithin{equation}{section}

\begin{document}


\begin{titlepage}

\vspace{-3cm}

\title{\hfill{\ns hep-th/0411071\\}
   \vskip 1cm
   {\Large Classification and Moduli K\"ahler Potentials of $G_{2}$ Manifolds}\\}
   \setcounter{footnote}{0}
\author{
{\ns\large Adam B Barrett\footnote{email: barrett@thphys.ox.ac.uk}
  \setcounter{footnote}{3}
  and Andr\'e Lukas\footnote{email: lukas@physics.ox.ac.uk}} \\[0.8em]
   {\it\ns The Rudolf Peierls Centre for Theoretical Physics,}\\
   {\it\ns University of Oxford}\\
   {\ns 1 Keble Road, Oxford OX1 3NP, UK} \\[0.2em] }
\date{}

\maketitle

\begin{abstract}\noindent
Compact manifolds of $G_2$ holonomy may be constructed by dividing a
seven-torus by some discrete symmetry group and then blowing up the
singularities of the resulting orbifold. We classify possible group
elements that may be used in this construction and use this
classification to find a set of possible orbifold groups. We then
derive the moduli K\"ahler potential for M-theory on the resulting
class of $G_2$ manifolds with blown up co-dimension four
singularities.
\end{abstract}

\thispagestyle{empty}

\end{titlepage}


\section{Introduction}
Compactification of M-theory on a seven-dimensional space of holonomy
$G_2$ provides a way of obtaining a four-dimensional theory with $N=1$
supersymmetry. It is well known that in order to obtain a realistic
compactified theory there must be singularities present on the $G_2$
manifold. Compactification of 11-dimensional supergravity on a smooth
$G_2$ manifold results in a theory containing only abelian gauge
multiplets and massless uncharged chiral multiplets \cite{Comp1, Comp2}.
To be more specific, within this framework, gauge fields descend from
dimensional reduction of the three-form field of 11-dimensional
supergravity. (There are no continuous symmetries of a manifold of
$G_2$ holonomy by which to obtain gauge fields.)  Chiral multiplets
arise from the metric moduli of the $G_2$ manifold and the associated
axions.

Non-abelian gauge symmetry arises when the $G_2$ manifold has
co-dimension four $A - D - E$ orbifold singularities. In addition,
chiral fermions, that are possibly charged under the gauge multiplets,
arise when the locus of such a singularity passes through an isolated
conical (co-dimension seven) singularity \cite{Atiyah}.

In the context of supergravity, one often work with smooth manifolds,
but in the present setting we should keep in mind that it is 
singular limits of the smooth $G_2$ manifolds which are ultimately
going to be interesting for phenomenology. (For the manifolds
constructed in this paper, the limit in which some or all blow-up moduli are
shrunk down to zero.)  \\

There has been much work done on the subject of M-theory on $G_2$
spaces, in which calculations are carried out for some generic,
possibly compact manifold of $G_2$ holonomy  
or else for some specific non-compact $G_2$ 
manifold~\cite{Atiyah}-\cite{Morris}. However, it is clear 
that potentially
realistic examples should rely on compact $G_2$ spaces whose
properties can be fairly well understood. An interesting set of
compact manifolds of $G_2$ holonomy, based on $G_2$ orbifolds, has
been constructed in \cite{Joyce}. It is an interesting task to pursue
this method to construct compact $G_2$ manifolds further and, hence, to
classify $G_2$ orbifolds and their associated, blown-up $G_2$
manifolds. In this paper we propose a method for such a
classification, and then construct an explicit class of manifolds,
many of which appear to be new.

Having obtained a class of manifolds of holonomy $G_2$ it would be
useful to be able to compare the four-dimensional effective theories
resulting from compactification on different members of the class. An
important ingredient for beginning such analysis is the
four-dimensional moduli K\"ahler potential. This has obvious
applications to various areas of study, for example, supersymmetry
breaking or the cosmological dynamics of moduli fields. For the $G_2$
manifold based on the simplest $G_2$ orbifold~\cite{Joyce}
$T^7/\mathbb{Z}_2^3$ the K\"ahler potential has been calculated in
Ref.~\cite{Lukas}. In this paper, we generalize this result by deriving
a formula for the moduli K\"ahler potential valid for the manifolds
of our classification.

There are classes of internal manifolds for which a general, explicit
formula for the moduli K\"ahler potential exists (at least at tree
level) in terms of certain topological data. For example, Calabi-Yau
three-folds have their moduli K\"ahler potential determined by a cubic
polynomial with coefficients given by their triple intersection
numbers \cite{Strominger, Candelas}. Key to this result is a quasi
topological relation between two-forms and their Hodge duals on the
Calabi-Yau space. There appears to be no analogue for three-forms on
$G_2$ manifolds. There do exist \cite{witten} abstract formulae for
the moduli K\"ahler metric in terms of harmonic three-forms on the
$G_2$ manifold, and for the K\"ahler potential in terms of the volume
of the $G_2$ manifold, but these cannot be evaluated generically for
all $G_2$ manifolds in an analogous way to Calabi-Yau K\"ahler moduli
spaces. Our approach to compute the K\"ahler potential is to
explicitly construct all required objects for all members of our class
of manifolds. Concretely, we construct a family of $G_2$ structures
$\varphi$ with small torsion, and the associated family of ``almost
Ricci-flat'' metrics $g$. These are used to calculate the volume and
the periods which, combined, lead to an explicit expression for the
K\"ahler potential.\\

Let us give a brief definition of what a $G_2$ manifold is so that we
can describe the general idea~\cite{Joyce} of how to construct compact
$G_2$ manifolds from $G_2$ orbifolds. A $G_2$ manifold is a
seven-dimensional Riemmanian manifold admitting a globally defined
torsion free $G_2$ structure \cite{Joyce}. A $G_2$ structure
is given by a three-form $\varphi$ which can be written locally as
\begin{eqnarray}
\varphi & = &
 \mathrm{d}x^1\wedge\mathrm{d}x^2\wedge\mathrm{d}x^3+
 \mathrm{d}x^1\wedge\mathrm{d}x^4\wedge\mathrm{d}x^5+
 \mathrm{d}x^1\wedge\mathrm{d}x^6\wedge\mathrm{d}x^7+
 \mathrm{d}x^2\wedge\mathrm{d}x^4\wedge\mathrm{d}x^6\nonumber \\& &
-\mathrm{d}x^2\wedge\mathrm{d}x^5\wedge\mathrm{d}x^7-
 \mathrm{d}x^3\wedge\mathrm{d}x^4\wedge\mathrm{d}x^7-
 \mathrm{d}x^3\wedge\mathrm{d}x^5\wedge\mathrm{d}x^6\; .
\label{structure}
\end{eqnarray}
The $G_2$ structure is torsion-free if $\varphi$ satisfies
$\mathrm{d}\varphi=\mathrm{d}\star\varphi=0$. A
$G_2$ manifold has holonomy $G_2$ if and only if its first fundamental
group is finite. Our starting point for constructing a compact
manifold of $G_2$ holonomy is an arbitrary seven-torus $T^7$. We then
take the quotient with respect to a finite group $\Gamma$ contained in
$G_2$, such that the resulting orbifold has finite first fundamental
group. We shall refer to $\Gamma$ as the orbifold group. The result is
a $G_2$ manifold with singularities at fixed loci of elements of
$\Gamma$. Smooth $G_2$ manifolds are then obtained by blowing up the
singularities. Loosely speaking, this involves removing a patch around
the singularity and replacing it with a smooth space of the same
symmetry. Note that, following this construction, the independent
moduli will come from torus radii and from the radii and orientation
of cycles associated with the blow-ups.

Let us also give a brief outline of what is involved in the
calculation of the moduli K\"ahler potential. On a $G_2$ manifold
$\mathcal{M}$, a given $G_2$ structure $\varphi$ induces a Riemannian
metric \cite{Lukas}. Ricci-flat deformations of the metric can be
described by the torsion-free deformations of $\varphi$, and hence by
the third cohomology $H^3(\mathcal{M},\mathbb{R})$. Consequently, the
number of independent metric moduli is given by the third Betti number
$b^3(\mathcal{M})$. To define these moduli explicitly, we introduce an
integral basis $\{C^I\}$ of three-cycles, and a dual basis
$\{\Phi_I\}$ of harmonic three forms satisfying
\begin{equation} \label{dual}
\int_{C^I}\Phi_J=\delta^I_J.
\end{equation}
Here $I,J,\ldots=1,\ldots,b^3(\mathcal{M})$. We can then expand $\varphi$ as 
\begin{equation}
\varphi=\sum_Ia^I\Phi_I,
\end{equation}
where the coefficients $a^I$ are precisely our metric moduli. Then, by
equation \eqref{dual}, the $a^I$ can be computed in terms of certain
underlying geometrical parameters (on which the $G_2$ structure $\varphi$
depends) by performing the period integrals
\begin{equation}
a^I=\int_{C^I}\varphi.
\end{equation}
In the four-dimensional effective theory the $a^I$ form the real
bosonic part of $b^3(\mathcal{M})$ chiral superfields $T^I$. (The
corresponding imaginary parts are axionic fields descending from
dimensional reduction of the three-form field of 11-dimensional
supergravity.) It is the K\"ahler potential for these fields $T^I$
which we wish to compute explicitly. Our strategy for doing this will
be to compute the volume of the manifold and to then use the formula
\begin{equation} \label{witten}
K=-3\ln\left(\frac{V}{2\pi^2}\right)
\end{equation}
from Ref.~\cite{witten}. With the result for the period integrals this
expression for $K$ can then be re-written in terms of the $a^I$ and,
hence, the superfields $T^I$.

Ideally one would like to perform the calculation using a torsion-free
$G_2$ structure. However, such torsion-free structures are not known
explicitly on compact $G_2$ manifolds. Instead, as in Ref.~\cite{Joyce}, we
write down explicit $G_2$ structures with small torsion and compute
the K\"ahler potential in a controlled approximation, knowing
that there exist ``nearby'' torsion-free $G_2$ structures.\\

The plan of the paper is as follows. In the next two sections we
describe the classification of orbifold-based compact $G_2$
manifolds. The classification shall be in terms of the orbifold group
of the manifold, and so in Section \ref{elements} we draw up a list of
symmetries of seven-tori that may be used as generators of orbifold
groups. For a symmetry $\alpha$ to be suitable for the orbifolding
there must exist a $G_2$ structure $\varphi$ on the torus that is
preserved by $\alpha$. Then in Section \ref{groups} we look at ways of
combining these symmetries to form orbifold groups that give orbifolds
of finite first fundamental group. We find a straightforward way of
checking when this condition is satisfied. A summary of the results is
as follows. There is only one possible abelian orbifold group,
$\mathbb{Z}_2^3$, with three or less generators. Further, we have been
looking for viable examples within the class of orbifold groups formed
by three or less generators with co-dimension four singularities
subject to an additional technical constraint on the allowed
underlying lattices. Within this class we have found all viable
examples consisting of ten distinct semi-direct product groups with
three generators as well as five exceptional cases built from three
generators with a more complicated algebra.

In Section \ref{manifolds} we give a description of a general $G_2$
manifold with blown up co-dimension four orbifold fixed points of type
$A$. We present a basis of its third homology, and write down formulae
for metrics and $G_2$ structures of small torsion. Then in Section
\ref{kaehler}, having all the machinery in place, we compute the
moduli K\"ahler potential for our class of $G_2$ manifolds, valid
for sufficiently small blow-up moduli.

To keep the main text more readable we have collected some of the
technical details in Appendices \ref{a} and \ref{c}. Appendix \ref{a}
contains some results on $G_2$ structures useful for the
classification of Sections \ref{elements} and \ref{groups}, whilst
Appendix \ref{c} has some of the details of how to blow-up
singularities and some calculations on the associated Gibbons-Hawking
spaces. Finally, Appendix \ref{b} contains a table listing the
possible orbifold group elements.


\section{Classification of Orbifold Group Elements} \label{elements}
The most general group element $\alpha$ that we consider acts on
seven-dimensional vectors $\boldsymbol{x}$ by $\alpha : \boldsymbol{x}
\mapsto A_{(\alpha)}\boldsymbol{x} + \boldsymbol{b}_{(\alpha)}$, where
$A_{(\alpha)}$ is an orthogonal matrix and $\boldsymbol{b}_\alpha$ is
a shift vector. We shall find all such $\alpha$ that give a consistent
orbifolding of some seven-torus, and that preserve some $G_{2}$
structure. Mathematically, by a seven-torus, we mean the fundamental
domain of $\mathbb{R}^7/\Lambda$, where $\Lambda$ is some seven-dimensional 
lattice. In the following we shall use bold Greek letters to denote
lattice vectors.

Let us begin by considering the orbifolding. In the following, bold
Greek letters denote lattice vectors. For consistency we require that
two points $\boldsymbol{x}$ and $\boldsymbol{y}$ in one unit cell are
equivalent under the orbifolding if and only if the corresponding two
points $\boldsymbol{x}+\boldsymbol{\lambda}$ and
$\boldsymbol{y}+\boldsymbol{\lambda}$ in another unit cell are
equivalent. Suppose
\begin{equation}
\boldsymbol{y}=  A_{(\alpha)}\boldsymbol{x} + \boldsymbol{b}_{(\alpha)}
                 + \boldsymbol{\mu}.
\end{equation}
Then there must exist a $ \boldsymbol{\nu}$ such that
\begin{equation}
\boldsymbol{y} + \boldsymbol{\lambda}=  A_{(\alpha)}(\boldsymbol{x}+
  \boldsymbol{\lambda}) + \boldsymbol{b}_{(\alpha)} + \boldsymbol{\nu}.
\end{equation}
It follows that $\alpha$ gives a consistent orbifolding precisely when
$A_{(\alpha)}$ takes lattice vectors to lattice vectors. Note that
there is no constraint on $ \boldsymbol{b}_{(\alpha)}$.

We are interested in classifying orthogonal matrices that preserve a
seven-dimensional lattice. A useful starting point is to find out the 
possible orders of such orthogonal group elements.  We first
quote a result from Ref.~\cite{Ono} in $n$-dimensions and then apply it to
$n=7$. Let $A\in O(n)$ be of order $m$. Then all its eigenvalues are
$m^{\mathrm{th}}$ roots of unity, and so we can choose an orthonormal
basis for $\mathbb{R}^{n}$ with respect to which $A$ takes the form
\begin{equation} 
A= \left( \begin{array}{cccc}
A_{1} & \, & \, & \,  \\
\, & A_{2} & \, & \,  \\
\, & \, & \ddots & \, \\
\, & \, & \, & A_{k}  \\
\end{array} \right) ,  
\end{equation}  
with $A_{j} \in O(d_{j})$, the eigenvalues of $A_{j}$ being primitive
$m_{j}^{\mathrm{th}}$ roots of unity, $m_{j}\lvert m$, and $m_{i} \neq
m_{j}$ for $i \neq j$. The result is that there exists a seven-dimensional 
lattice preserved by $A$ if and only if we can write each $d_j$ in the form
\begin{equation} \label{euler}
d_{j}=n_{j} \varphi (m_{j}),
\end{equation}
where $n_{j} \in \mathbb{N}$ and $\varphi (m)$ is Euler's
function, the number of integers less than $m$ that are prime to
$m$. Furthermore, for an $A$ that satisfies \eqref{euler}, each primitive 
$m_{j}^{\mathrm{th}}$ root of unity is
an eigenvalue of $A_{j}$ with geometric multiplicity precisely
$n_{j}$.

We now consider the values $m_{j}$ is allowed to take when
$n=7$. Since $d_{j}\le 7$, $A_{j}$ can only be a constituent block of
$A$ if $\varphi(m_{j}) \le 7$. There is a formula from number theory
\begin{equation} \label{5}
\varphi (a) = \prod_{i} (p_{i}-1)p_{i}^{r_{i}-1},
\end{equation} 
where now $a=\prod p_{i}^{r_{i}}, r_{i}\in \mathbb{N}$ is the prime
decomposition of $a$. It is then straightforward to show from
Eq.~\eqref{5} that the allowed values of $m_{j}$ are $m_{j}=1, 2, \ldots,
10, 12, 14, 18$. It is also easy to see that this result is the same
as for $n=6$, a fact which we will use, since the $n=6$ case is the
relevant one for classifying orbifold-based Calabi-Yau spaces.

We now look for conditions on $A$ to belong in $G_{2}$. According to
\cite{Joyce} we must have $A\in SO(7)$. It therefore has eigenvalues 1
and complex conjugate pairs of modulus one. Hence we can write $A$ in
the canonical form
\begin{equation}
A= \left( \begin{array}{cccc}
1 & \, & \, & \,  \\
\, & R(\theta_{1}) & \, & \,  \\
\, & \, & R(\theta_{2}) & \, \\
\, & \, & \, & R(\theta_{3})  \\
\end{array} \right) ,  
\end{equation} 
where
\begin{equation} \label{theta}
 R(\theta_{i})= \left( \begin{array}{cc}
\cos\theta_{i} & -\sin\theta_{i} \\
\sin\theta_{i} & \cos\theta_{i} \\
\end{array} \right).
\end{equation}
Accordingly, $A$ decomposes as 
\begin{equation} \label{decomp}
A=1 \oplus A^{\prime},
\end{equation}
where $A^{\prime} \in SO(6)$.

As an aside, let us just mention that naively one may have first
derived the decomposition \eqref{decomp} and then decided immediately
that this implies that the set of possible orders of $A$ is identical
to the set of possible orders of symmetries of six-dimensional
lattices. Although this does indeed turn out to be the case, it is not
obvious that the $A$ of \eqref{decomp} preserves a seven-dimensional
lattice if and only if the corresponding $A^{\prime}$ preserves a
six-dimensional lattice. Rather, this can be shown by
applying Eq.~\eqref{euler} which leads to the same
allowed values of $m_j$ for the cases $n=6$ and $n=7$.

Now $A\in G_{2}$ if and only if it leaves a $G_{2}$ structure
invariant. In other words $A$ must leave $\varphi$ defined by equation
\eqref{structure} invariant, or else there must exist an $O(7)$
transformation taking $\varphi$ to a three-form $\tilde{\varphi}$ that
\textit{is} left invariant by $A$. It is convenient to recast
$\varphi$ in complex form by taking
\begin{equation} \label{complex}
x_{0}=x_{1},\:
z_{1}=\frac{1}{\sqrt{2}}(x_{2}+ix_{3}),\:
z_{2}=\frac{1}{\sqrt{2}}(x_{4}+ix_{5}),\:
z_{3}=\frac{1}{\sqrt{2}}(x_{6}+ix_{7}),
\end{equation}
to obtain
\begin{eqnarray} \label{form}
\varphi & = & \mathrm{d}x_{0}\wedge i(\mathrm{d}z_{1}\wedge\mathrm{d}\bar{z}_{1}+
\mathrm{d}z_{2}\wedge\mathrm{d}\bar{z}_{2}+\mathrm{d}z_{3}\wedge\mathrm{d}\bar{z}_{3})
+\sqrt{2}\mathrm{d}z_{1}\wedge\mathrm{d}z_{2}\wedge\mathrm{d}z_{3} \nonumber\\
& & +\sqrt{2}\mathrm{d}\bar{z}_{1}\wedge\mathrm{d}\bar{z}_{2}
    \wedge\mathrm{d}\bar{z}_{3}.
\end{eqnarray}
We can then see by inspection that $A$ preserves $\varphi$ if and only if
\begin{equation}
\theta_{1}+\theta_{2}+\theta_{3}=0 \: \mathrm{mod} \: 2\pi .
\end{equation}
The following is also easily verified. Under a transformation only
containing reflections in coordinate axes, asking for the resulting
$\tilde{\varphi}$ to be left invariant by $A$ imposes one of the
following four conditions:
\begin{eqnarray}
 \theta_{1}+\theta_{2}+\theta_{3} & = & 0 \: \mathrm{mod} \: 2\pi, \label{one}\\
  -\theta_{1}+\theta_{2}+\theta_{3} & = & 0 \: \mathrm{mod} \: 2\pi, \label{two}\\
  \theta_{1}-\theta_{2}+\theta_{3} & = & 0 \: \mathrm{mod} \: 2\pi, \label{three}\\
  \theta_{1}+\theta_{2}-\theta_{3} & = & 0 \: \mathrm{mod} \: 2\pi. \label{four}
\end{eqnarray}
Hence that $A$ satisfies one of \eqref{one}, \eqref{two},
\eqref{three} or \eqref{four} is sufficient for $A\in G_{2}$. It turns
out that this is also necessary. The proof of this is somewhat
technical, and an outline of it is given in Appendix \ref{a}. It is
useful to note that the condition we have on $A$ to belong in $G_{2}$
is precisely the condition \cite{Bailin} on the $A^{\prime}$ of
\eqref{decomp} to belong in $SU(3)$ under some embedding of $SU(3)$ in
$SO(6)$.

By combining the above results we see that the classification is in
one-to-one correspondence with that of the possible orbifold group
elements of a Calabi-Yau space, which is given, for example, in
Refs.~\cite{Bailin} and \cite{Dixon}. The table in Appendix \ref{b} gives
this classification in terms of the rotation angels $\theta_{i}$.

\section{Classification of Orbifold Groups} \label{groups}
Having obtained a class of possible generators, we now wish to find a
class of discrete symmetry groups from which compact manifolds of
$G_2$ holonomy may be constructed. Let us state the conditions for
$\Gamma$ to be a suitable orbifold group. There must exist both a
seven-dimensional lattice $\Lambda$ and a $G_2$ structure
$\varphi$ that are preserved by $\Gamma$, and the first fundamental
group $\pi_1$ of $(\mathbb{R}^7/\Lambda)/\Gamma$ must be finite.

It is useful to translate the condition on $\pi_1$ into an equivalent
condition that is more readily checked. An equivalent condition is
that there exist no non-zero vectors $\boldsymbol{n}$ with the
property that $A_{(\alpha)}\boldsymbol{n}=\boldsymbol{n}$ for each
$\alpha\in\Gamma$. That this condition is sufficient for $\pi_1$ to be
finite is shown in Ref. \cite{Joyce}. That this is necessary is
demonstrated below.

Let $\{\boldsymbol{\lambda}_{j}\}$ be a basis of lattice vectors and
write $\boldsymbol{n}$ in the form
\begin{equation} \label{nexp}
\boldsymbol{n}=\sum_{j}n_{j}\boldsymbol{\lambda}_{j}.
\end{equation}
Let $\{\alpha_{1},\ldots,\alpha_{k}\}$ be the generators of
$\Gamma$. Then applying $\alpha_{l}$ to both sides of \eqref{nexp} we
have
\begin{equation} \label{nexp2}
\boldsymbol{n}=\sum_{j,i}n_{j}a_{ji}^{(l)}\boldsymbol{\lambda}_{i},
\end{equation}
where the $a_{ji}^{(l)}$ are matrices with integer coefficients. From
\eqref{nexp} and \eqref{nexp2} we obtain
\begin{equation} \label{system}
\sum_{j}n_{j}b_{ji}^{(l)}=0, \: \: \forall i,l,
\end{equation}
where $b_{ji}^{(l)}=a_{ji}^{(l)}-\delta_{ji}$. Now since, by
assumption, there exist non-zero solutions to \eqref{system}, in
constructing a particular solution we may choose the value of at least
one of the $n_i$'s. Let us set the value of this $n_i$ to unity. Now
consider another $n_i$. If it is free then let us set it also to
unity. If it is constrained then it must be a linear function of other
$n_i$'s with rational coefficients. We have hence constructed a
solution to \eqref{system} with each $n_i$ rational. Now, for our
solution, write $n_i$ in the form
\begin{equation}
n_i=\frac{p_i}{q_i},
\end{equation}
with $p_i$ and $q_i$ integers. Then
$\mathrm{lcm}\{q_i\}\boldsymbol{n}$ is a lattice vector (where lcm
stands for ``lowest common multiple'') and so the path that is a
straight line from the origin of the orbifold to
$w\,\mathrm{lcm}\{q_i\}\boldsymbol{n}$, with $w$ an integer, is a path
of winding number $w$. This establishes the result.  \\

Given the above result, it is clear from \eqref{decomp} that a group
$\Gamma$ must contain more than one generator if the resulting
orbifold $T^{7}/\Gamma$ is to be of holonomy $G_{2}$. We now attempt
to construct abelian groups of the form
$\mathbb{Z}_m\times\mathbb{Z}_n$ for which the corresponding orbifold
has holonomy $G_2$. For now we take the generator of the
$\mathbb{Z}_n$ symmetry to be a straightforward rotation
\begin{equation} \label{R}
R= \left( \begin{array}{cccc}
1 & \, & \, & \,  \\
\, & R(\theta_{1}) & \, & \,  \\
\, & \, & R(\theta_{2}) & \, \\
\, & \, & \, & R(\theta_{3})  \\
\end{array} \right) ,  
\end{equation}
with $R(\theta_i)$ as in \eqref{theta} and
$(\theta_1,\theta_2,\theta_3)$ one of the triples of the table in
Appendix A.  We look for a second symmetry, also a pure rotation, with
corresponding matrix $P$ commuting with $R$ such that the group
generated by $P$ and $R$ is a suitable orbifold group. To find the
constraints on $P$ coming from commutativity we apply a generalization
of Schur's Lemma, as follows.

Write the reducible representation $R$ of the group $G$ as
$R=n_1R_1\oplus\cdots\oplus n_rR_r$, where the $R_i$ are irreducible
representations of $G$ of dimension $d_i$ and the integers $n_i$
indicate how often each $R_i$ appears in $R$. Then a matrix $P$ with
$[P,R(g)]=0$ for all $g\in G$ has the general form
\begin{equation} \label{Schur}
P=P_1\otimes \boldsymbol{1}_{d_1}\oplus\cdots\oplus P_r\otimes \boldsymbol{1}_{d_r},
\end{equation}
where the $P_i$ are $n_i\times n_i$ matrices.

The table in Appendix \ref{b} lists the $n_i$ and $d_i$ for each $R$
we are considering. We can therefore simply go through this table and,
for each $R$, see if there are any $P$s that are suitable for our
construction. It turns out that there are in fact no suitable $P$s for
any of the $R$s. In fact every case fails because $\pi_1$ is not finite. Below is one example to provide an illustration.

There is the possibility that $R$ represents a $\mathbb{Z}_2$ symmetry of the lattice and is given by
\begin{equation}
R  = \mathrm{diag}(1,1,1,-1,-1,-1,-1).
\end{equation}
Then we have $n_1=3$, $d_1=1$ and $n_2=4$, $d_2=1$. Applying the lemma \eqref{Schur}, $P$ must take the form
\begin{equation}
P= \left( \begin{array}{cc}
P_{1}^{3\times 3} & \,  \\
\, &  P_{2}^{4\times 4} \\
\end{array} \right).
\end{equation}
Now (see Appendix \ref{a}), any $G_2$ structure preserved by $R$ has
$\lvert\varphi_{123}\rvert=1$. Under $P$,
$\varphi_{123}\mapsto\mathrm{det}(P_{1}^{3\times 3})\varphi_{123}$ and
so we must have $\mathrm{det}(P_{1}^{3\times 3})=1$. Hence
$P_{1}^{3\times 3}\in SO(3)$, and leaves at least one direction
fixed. But $R$ fixes this direction too, since it fixes all of the 1,
2 and 3 directions. We therefore rule out this case since we can not
render $\pi_1$ finite by this construction.  \\

Let us now attempt to construct an orbifold group of the form
$\mathbb{Z}_m\times\mathbb{Z}_n\times\mathbb{Z}_p$. As above, we let
$R$ generate $\mathbb{Z}_p$, and use the same method as above to find
the possibilities for the other generators $P$ and $Q$. In looking for
$P$, most possibilities are still ruled out, but we can now relax the
condition that there are no non-zero fixed vectors of the group
generated by $R$ and $P$. There are then three cases we need to
consider. Firstly,
$\frac{1}{2\pi}(\theta_{1},\theta_{2},\theta_{3})=(\frac{1}{4},\frac{1}{4},
\frac{1}{2})$. In
this case, we find that $P$ must take the form
\begin{equation}
P =  (-1)\oplus \left( \begin{array}{cc}
\cos\phi_{1} & \sin\phi_{1} \\
\sin\phi_{1} & -\cos\phi_{1} \\
\end{array} \right) \oplus
 \left( \begin{array}{cc}\cos\phi_{2} & \sin\phi_{2} \\
\sin\phi_{2} & -\cos\phi_{2} \\
\end{array} \right) \oplus 
 \left( \begin{array}{cc} \cos\phi_{3} & \sin\phi_{3} \\
\sin\phi_{3} & -\cos\phi_{3} \\
\end{array} \right). 
\end{equation} 
Here the $2\times 2$ blocks each represent the most general element of
$O(2)-SO(2)$. It is easily verified that such a matrix can not commute
with $R$, thus ruling out this case. (Note that Eq.~\eqref{Schur} was a
necessary but not sufficient condition for commutativity.) The second
case is when $n_1=3$, $d_1=1$ and $n_2=2$, $d_2=2$, for which $p=3, 4$
or $6$. On checking the possibilities for this case we find that there
is no way of forming a group with all the correct properties. We are
left with one remaining case, the case with $p=2$, and this leads to
the group $\mathbb{Z}_2^{3}$. For this case, the matrices $P$, $Q$ and
$R$ are given essentially uniquely by
\begin{equation} \label{Z2}
R  = \mathrm{diag}(1,1,1,-1,-1,-1,-1),
\end{equation}
\begin{equation} \label{Z3}
P  =  \mathrm{diag}(1,-1,-1,-1,-1,1,1),
\end{equation}
\begin{equation} \label{Z4}
Q =  \mathrm{diag}(-1,-1,1,1,-1,-1,1).
\end{equation}
\\

The next step is to consider not just pure rotations, but to now allow
the group elements to contain translations as well. Let us derive a
condition for commutativity. Let two generators of an abelian orbifold
group be given by
\begin{equation}
\alpha : \boldsymbol{x} \mapsto A\boldsymbol{x} + \boldsymbol{a},
\end{equation}
\begin{equation}
\beta : \boldsymbol{x} \mapsto B\boldsymbol{x} + \boldsymbol{b}.
\end{equation}
Then commutativity requires us to be able to write
\begin{equation}
(\alpha\circ\beta)\boldsymbol{x}=(\beta\circ\alpha)\boldsymbol{x}+\sum_jn_j\boldsymbol{\lambda}_j,
\end{equation}
where the $\boldsymbol{\lambda}_j$ form a basis of lattice vectors and the $n_j$ are integers. This gives
\begin{equation}
[A,B]\boldsymbol{x}+A\boldsymbol{b}-B\boldsymbol{a}+\boldsymbol{a}-\boldsymbol{b}=\sum_jn_j\boldsymbol{\lambda}_j.
\end{equation}
Since $\boldsymbol{x}$ may vary continuously we see that we must still have $[A,B]=0$ and then we are left with
\begin{equation} \label{commute}
(A-I)\boldsymbol{b}-(B-I)\boldsymbol{a}=\sum_jn_j\boldsymbol{\lambda}_j.
\end{equation}
We are now able to write down the most general abelian orbifold group,
with at most three generators, from which a $G_2$ manifold may be
constructed. We apply the constraint \eqref{commute} to any
translations added to the matrix transformations of equations
\eqref{Z2}-\eqref{Z4}. The result is that the most general set of
generators act as follows on a vector $\boldsymbol{x}=(x_1, x_2, x_3,
x_4, x_5, x_6, x_7)$ of the standard seven-torus.
\begin{equation}
\alpha : \boldsymbol{x}\mapsto \bigg(x_1+\frac{m_1}{2},x_2+\frac{m_2}{2}, x_3+\frac{m_3}{2}, -x_4+a_4,-x_5+a_5,-x_6+a_6,-x_7+a_7\bigg),
\end{equation}
\begin{equation}
\beta : \boldsymbol{x}\mapsto \bigg(x_1+\frac{n_1}{2},-x_2+b_2, -x_3+b_3,  -x_4+b_4,-x_5+b_5,x_6+\frac{n_6}{2},x_7+\frac{n_7}{2}\bigg),
\end{equation}
\begin{equation}
\gamma : \boldsymbol{x}\mapsto \bigg(-x_1+c_1,-x_2+c_2, x_3+\frac{p_3}{2}, x_4+\frac{p_4}{2}, -x_5+c_5,-x_6+c_6,x_7+\frac{p_7}{2}\bigg),
\end{equation}
where the $m_i$, $n_i$ and $p_i$ are integers and the $a_i$, $b_i$ and $c_i$ are unconstrained reals. The group generated is always $\mathbb{Z}_2^{3}$.
\\

Our objective was to find a class of orbifold groups and to achieve
this, it appears that commutativity is not the most suitable
constraint to impose, in spite of the systematic approach it gave
us. Since well-defined procedures to describe the metric on the
blow-ups are available for the cases with co-dimension four fixed
loci, we now focus on orbifold groups that only lead to these. We thus
restrict attention to generators that leave three directions of the
torus invariant, namely those whose rotation part is one of
$\mathbb{Z}_2$, $\mathbb{Z}_3$, $\mathbb{Z}_4$ or $\mathbb{Z}_6$ from
the table in Appendix \ref{b}. In fact, for this section we shall
consider pure rotations only. We show that a simple constraint on the
lattice itself enables us to come up with a substantial class of
possible orbifold groups. Let us insist that, for each generator
$\alpha$ of the orbifold group, there exists a partition of our basis
of lattice vectors into three sets, spanning the spaces $U$, $V$ and
$W$, of dimension 3, 2 and 2 respectively such that
\begin{equation} \label{latt1}
\Lambda =U\perp V\perp W,
\end{equation}
\begin{equation} \label{latt2}
\alpha U = U,\;\; \alpha V = V, \; \; \alpha W = W,
\end{equation}
\begin{equation}\label{latt3}
\alpha \lvert_U = \iota.
\end{equation}
This seems a sensible condition to impose, since it makes it easy to picture 
how the orbifold group acts on the lattice. Basically, each generator rotates two two-dimensional sub-lattices.

The classification of orbifold groups, subject to the above constraints and 
containing three or fewer generators, now goes as follows. We take as the 
first generator the
matrix $R$, in the canonical form of \eqref{R} with $\theta_1=0$ and
$\theta_2=-\theta_3=2\pi / N$, where $N=2,$ 3, 4, or 6. We then use
coordinate freedom to choose a $G_2$ structure $\varphi$ that is the
standard one of Eq.~\eqref{structure} up to possible sign differences (see
Appendix \ref{a}).

We are then able to derive the other possible generators of the
orbifold group that are distinct up to redefinitions of the
coordinates. First we narrow the possibilities using the constraint of
$G_2$ structure preservation. In particular this imposes the condition
that the three fixed directions must correspond precisely to one of
the seven terms in the $G_2$ structure (see Appendix \ref{a}). Having
done this we look for a preserved lattice. It is straightforward to go
through all possibilities having imposed \eqref{latt1}, \eqref{latt2}
and \eqref{latt3}. We find the following distinct generators. Firstly
matrices in the canonical form of \eqref{R} with
$\theta_1=-\theta_2=2\pi / M$ and $\theta_3=0$, with $M$ dividing $N$
or vice-versa (with the exception $M=2$, $N=3$). Secondly
$\mathbb{Z}_2$ symmetries not in canonical form, for example
\begin{equation} \label{Q}
Q_0 =  \mathrm{diag}(-1,1,-1,1,-1,1,-1),
\end{equation}
and then, only for the cases $N=2$ or 4, $\mathbb{Z}_4$ symmetries not
in canonical form, for example
\begin{equation}
Q_1 : (x_1,x_2,x_3,x_4,x_5,x_6,x_7) \mapsto (x_7,x_2,-x_5,x_4,x_3,x_6,-x_1).
\end{equation}

We can now go about combining these generators together. It is easy to see that, as when we were considering abelian groups, there are no orbifold groups built from just two generators that result in orbifolds with finite first fundamental group. Our class therefore consists solely of orbifold groups containing three generators. The main sub-class has generators $P$ and
$R$ in the canonical form of \eqref{R}, with $P$ having
$\theta_1=-\theta_2=2\pi / M$ and $\theta_3=0$ and $R$ having
$\theta_1=0$ and $\theta_2=-\theta_3=2\pi / N$, and $Q_0$ as given in
\eqref{Q}. This class contains the following orbifold groups:
\begin{equation}
\begin{array}{l}
\mathbb{Z}_2 \times \mathbb{Z}_2 \times \mathbb{Z}_2, \\
\mathbb{Z}_2 \ltimes \left( \mathbb{Z}_2 \times \mathbb{Z}_3 \right), \\
\mathbb{Z}_2 \ltimes \left( \mathbb{Z}_2 \times \mathbb{Z}_4 \right), \\
\mathbb{Z}_2 \ltimes \left( \mathbb{Z}_2 \times \mathbb{Z}_6 \right) ,\\
\mathbb{Z}_2 \ltimes \left( \mathbb{Z}_3 \times \mathbb{Z}_3 \right) ,\\
\mathbb{Z}_2 \ltimes \left( \mathbb{Z}_3 \times \mathbb{Z}_6 \right) ,\\
\mathbb{Z}_2 \ltimes \left( \mathbb{Z}_4 \times \mathbb{Z}_4 \right) ,\\
\mathbb{Z}_2 \ltimes \left( \mathbb{Z}_6 \times \mathbb{Z}_6 \right) .\\
\end{array}
\end{equation}
The semi-direct product notation is defined by writing $G\ltimes H$ if
$G$ and $H$ are abelian and $[g,h]\in H$ for any $g\in G$ and $h\in
H$. In fact, within our class of orbifold groups, the stronger condition 
$[g,h]=h^2$ will be satisfied in each case of a semi-direct product. Note 
that since we should really be thinking of orbifold group elements as 
abstract group elements as opposed to matrices, the commutator is defined by $[g,h]=g^{-1}h^{-1}gh$. Some 
other semi-direct products are attained by using $R$ as above
and then taking $Q_0$ and $\mathrm{diag}(-1,-1,1,1,-1,-1,1)$ as the
other generators to obtain the following groups:
\begin{equation}
\mathbb{Z}_2^2 \ltimes \mathbb{Z}_N, \; \; N=3, \, 4 \,\, \mathrm{or} \,\, 6.
\end{equation}

There are also some exceptional cases that contain a $\mathbb{Z}_4$
not in canonical form. These exceptional orbifold groups will contain
only $\mathbb{Z}_2$ and $\mathbb{Z}_4$ symmetries. We make the
observations that $\mathbb{Z}_2$ and $\mathbb{Z}_4$ symmetries either
commute or give semi-direct products and that two $\mathbb{Z}_4$
symmetries $A$ and $B$ either commute or have the relation
$A^2BA^2=B^{-1}$. It is then straightforward to find all possible
group algebras for our exceptional orbifold groups. Going through the
conditions for a $G_2$ orbifold, we find some of these algebras can be
realised and some can not. Those that can are constructed as
follows. Take $P$ and $R$ as before with $M=2$ and $N=4$ and use
either $Q_1$ from above or $Q_2$ given by
\begin{equation}
Q_2 : (x_1,x_2,x_3,x_4,x_5,x_6,x_7) \mapsto (x_3,x_2,-x_1,x_4,-x_7,x_6,x_5).
\end{equation}
These lead to the respective groups
\begin{equation}
\mathbb{E}_1 =: \langle P,\, Q_1,\, R\, \lvert\, P^2=1,\, Q_1^4=1,\, R^4=1,\, [P,Q_1]=1,\, [P,R]=1,\, Q_1^2RQ_1^2=R^{-1}  \rangle,
\end{equation}
and
\begin{equation}
\mathbb{E}_2 =: \langle P,\, Q_2,\, R\, \lvert\, P^2=1,\, Q_2^4=1,\, R^4=1,\, [P,Q_2]=Q_2^2,\, [P,R]=1,\, Q_2^2RQ_2^2=R^{-1}  \rangle.
\end{equation}
Then three more possibilities come about from using $R$ with $N=4$, $Q_2$ and one of the $P_i$ given below. 
\begin{equation}
P_1 : (x_1,x_2,x_3,x_4,x_5,x_6,x_7) \mapsto (-x_1,-x_2,x_3,-x_4,x_5,x_6,-x_7),
\end{equation}
\begin{equation}
P_2 : (x_1,x_2,x_3,x_4,x_5,x_6,x_7) \mapsto (x_1,-x_3,x_2,x_5,-x_4,x_6,x_7),
\end{equation}
\begin{equation}
P_3 : (x_1,x_2,x_3,x_4,x_5,x_6,x_7) \mapsto (x_1,-x_5,x_4,-x_3,x_2,x_6,x_7).
\end{equation}
The groups we obtain can be described respectively by
\begin{equation}
\mathbb{E}_3 =: \langle P_1,\, Q_2,\, R\, \lvert\, P_1^2=1,\, Q_2^4=1,\, R^4=1,\, [P_1,Q_2]=Q_2^2,\, [P_1,R]=R^2,\, Q_2^2RQ_2^2=R^{-1}  \rangle,
\end{equation}
\begin{equation}
\mathbb{E}_4 =: \langle P_2,\, Q_2,\, R\, \lvert\, P_2^4=1,\, Q_2^4=1,\, R^4=1,\, P_2^2Q_2P_2^2=Q_2^{-1},\, [P_2,R]=1,\, Q_2^2RQ_2^2=R^{-1}  \rangle,
\end{equation}
\begin{equation}
\mathbb{E}_5 =: \langle P_3,\, Q_2,\, R\, \lvert\, P_3^4=1,\, Q_2^4=1,\, R^4=1,\, P_3^2Q_2P_3^2=Q_2^{-1},\, P_3^2RP_3^2=R^{-1},\, Q_2^2RQ_2^2=R^{-1}  \rangle.
\end{equation}
\\

It is worth briefly summing up what we have found. We have found a
class of sixteen distinct groups, each composed of three pure
rotations, that may be used as orbifold groups to construct compact
manifolds of $G_2$ holonomy. If two simple constraints are imposed on
the manifolds, we have the complete class of such orbifold groups. The
first constraint states that the manifold is to have only co-dimension
four fixed points. The second constraint is on the lattice from which
the manifold is constructed. It states that the lattice must decompose
into an orthogonal sum of smaller lattices, with three constituents of
the sum being simple two-dimensional lattices, and the final
constituent being the trivial one-dimensional lattice. Furthermore the
action of each orbifold group generator must be to simply rotate two
two-dimensional sub-lattices.

There are two obvious ways of extending the classification we have
obtained. One is to remove the restriction on the lattice. Perhaps
this would give rise to some more complicated examples, in which
lattice vectors do not lie precisely in the planes of rotation of the
orbifold group generators. A second method would be to allow
co-dimension six fixed points and thus allow any of the generators
listed in Appendix \ref{b}.

\section{Description of the Manifolds and their $G_2$ Structures} \label{manifolds}
In this section we describe in some detail the manifolds for which we
shall compute the moduli K\"ahler potential. We take $\mathcal{M}$ to
be a general smooth $G_2$ manifold, constructed from an orbifold
$\mathcal{O}=T^7/\Gamma$ with co-dimension four fixed points. We assume that
points on the torus that are fixed by one generator of the orbifold
group are not fixed by other generators. Given an orbifold group, this
can always be arranged by incorporating appropriate translations into
the generators, and thus all of our previously found examples are relevant. 
Under this assumption we have a well-defined blow-up
procedure.

Let us introduce some notation. We use the index $\tau$ to label the 
generators of the orbifold group $\Gamma$, and write $N_\tau$ for the order 
of the generator $\alpha_\tau$. For $\Gamma$ to only 
have co-dimension 
four fixed points the possible values of $N_\tau$
are 2, 3, 4 and 6 (see Appendix \ref{b}). Each generator will have a
certain number $M_\tau$ of fixed points associated with it. A singular
point on $\mathcal{O}$ is therefore labelled by a pair $(\tau,n)$,
where $n=1,\ldots,M_\tau$. 

Near a singular point $\mathcal{O}$ takes the approximate form
$T^3_{(\tau,n)}\times\mathbb{C}^2/\mathbb{Z}_{N_{\tau}}$, where
$T^3_{(\tau,n)}$ is a three-torus. Blowing up the singularity involves
the following. One firstly removes a four-dimensional ball centred
around the singularity times the associated fixed three-torus
$T^3_{(\tau,n)}$. Secondly one replaces the resulting hole by
$T^3_{(\tau,n)}\times U_{(\tau,n)}$, where $U_{(\tau,n)}$ is the
blow-up of $\mathbb{C}^2/\mathbb{Z}_{N_{\tau}}$ as discussed in
Appendix \ref{c}.  \\

Before giving a more detailed description of what $\mathcal{M}$ looks
like, let us present a basis of three-cycles. This will be needed to
compute the periods and hence the moduli in terms of underlying
geometrical parameters. Localised on the blow-up labelled by
$(\tau,n)$ there are $3(N_\tau-1)$ three-cycles. These are formed by
taking the Cartesian product of one of the $(N_\tau-1)$ two-cycles on
$U_{(\tau,n)}$ with one of the three one-cycles on $T^3_{(\tau,n)}$.
Let us label these three-cycles by $C(\tau,n,a,i)$, where $a$ labels
the direction on $T^3_{(\tau,n)}$ and $i$ labels the two-cycles of
$U_{(\tau,n)}$. On the bulk, that is the remaining parts of the torus,
we can define three-cycles by setting four of the coordinates $x^A$ to
constants (chosen so there is no intersection with any of the
blow-ups). The number of these that fall into distinct homology
classes is then given by the number of independent terms in the $G_2$
structure on the bulk. Let us explain this statement. The bulk $G_2$
structure can always be chosen so as to contain the seven terms of the
standard $G_2$ structure \eqref{structure}, with positive coefficients
multiplying them. If we write $\mathcal{R}^A$ for the coefficient in
front of the $A^{\mathrm{th}}$ term in Eq.~\eqref{structure}, then by
the number of independent terms we mean the number of $\mathcal{R}^A$s
that are not constrained by the orbifolding. We then write $C^A$ for
the cycle obtained by setting the four coordinates on which the
$A^{\mathrm{th}}$ term in \eqref{structure} does not depend to
constants, for example,
\begin{equation}
C^1=\{x^4,x^5,x^6,x^7=\mathrm{const}\}.
\end{equation}
A pair of $C^A$s for which the corresponding $\mathcal{R}^A$s are
independent then belong to distinct homology classes. There is
therefore some subset $\mathcal{C}$ of $\{C^A\}$ such that the
collection $\{\mathcal{C},C(\tau,n,a,i)\}$ provides a basis for
$H_3(\mathcal{M},\mathbb{Z})$. We deduce the following formula for the
third Betti number of $\mathcal{M}$:
\begin{equation}
b^3(\mathcal{M})=b(\Gamma)+\sum_\tau M_\tau \cdot 3(N_\tau-1),
\end{equation}
where $b(\Gamma)$ is the number of bulk three-cycles, a positive
integer less or equal than seven, and dependent on the orbifold group
$\Gamma$. For the class of orbifold groups obtained in the previous
section $b(\Gamma)$ takes values as given in Table~1. A description of
the derivation is given in the discussion below on constructing the
bulk $G_2$ structure.  \\

\begin{table}
\begin{center}
\begin{tabular}{c|c}
$\Gamma$ & $b(\Gamma)$ \\
\hline
$\mathbb{Z}_2\times\mathbb{Z}_2\times\mathbb{Z}_2$ & 7 \\
$\mathbb{Z}_2 \ltimes \left( \mathbb{Z}_2 \times \mathbb{Z}_3 \right)$ & 5 \\
$\mathbb{Z}_2 \ltimes \left( \mathbb{Z}_2 \times \mathbb{Z}_4 \right)$ & 5 \\
$\mathbb{Z}_2 \ltimes \left( \mathbb{Z}_2 \times \mathbb{Z}_6 \right)$ & 5\\
$\mathbb{Z}_2 \ltimes \left( \mathbb{Z}_3 \times \mathbb{Z}_3 \right)$ & 4\\
$\mathbb{Z}_2 \ltimes \left( \mathbb{Z}_3 \times \mathbb{Z}_6 \right)$ & 4\\
$\mathbb{Z}_2 \ltimes \left( \mathbb{Z}_4 \times \mathbb{Z}_4 \right)$ & 4\\
$\mathbb{Z}_2 \ltimes \left( \mathbb{Z}_6 \times \mathbb{Z}_6 \right)$ & 4\\
$\mathbb{Z}_2^2 \ltimes \mathbb{Z}_3$ & 5\\
$\mathbb{Z}_2^2 \ltimes \mathbb{Z}_4$ & 5\\
$\mathbb{Z}_2^2 \ltimes \mathbb{Z}_6$ & 5\\
$\mathbb{E}_1$ & 3\\
$\mathbb{E}_2$ & 3\\
$\mathbb{E}_3$ & 3\\
$\mathbb{E}_4$ & 2\\
$\mathbb{E}_5$ & 1\\
\end{tabular}
\end{center}
\caption{Bulk third Betti numbers of the orbifold groups}
\end{table}

We now discuss the geometrical structure of $\mathcal{M}$ in more
detail, focusing in particular on the blow-up regions. Metrics and
$G_2$ structures will be presented on each region of
$\mathcal{M}$. Let us begin with the bulk, which is the
straightforward part. Assuming that, for a constant metric, only the
diagonal components survive the orbifolding, which is certainly the
case for the explicit examples constructed in Section \ref{groups},
\begin{equation} \label{bulkmetric}
\mathrm{d}s^2=\sum_{A=1}^7(R^A\mathrm{d}x^A)^2.
\end{equation}
Here the $R^A$ are precisely the seven radii of the torus. Under a
suitable choice of coordinates the $G_2$ structure is obtained from
the flat $G_2$ structure \eqref{structure} by rescaling $x^A\to
R^Ax^A$, leading to
\begin{eqnarray} \label{structure3}
\varphi & = & R^1R^2R^3\mathrm{d}x^1\wedge\mathrm{d}x^2\wedge\mathrm{d}x^3+R^1R^4R^5\mathrm{d}x^1\wedge\mathrm{d}x^4\wedge\mathrm{d}x^5+R^1R^6R^7\mathrm{d}x^1\wedge\mathrm{d}x^6\wedge\mathrm{d}x^7 \nonumber \\ & &+R^2R^4R^6\mathrm{d}x^2\wedge\mathrm{d}x^4\wedge\mathrm{d}x^6 -R^2R^5R^7\mathrm{d}x^2\wedge\mathrm{d}x^5\wedge\mathrm{d}x^7-R^3R^4R^7\mathrm{d}x^3\wedge\mathrm{d}x^4\wedge\mathrm{d}x^7 \nonumber \\ & & -R^3R^5R^6\mathrm{d}x^3\wedge\mathrm{d}x^5\wedge\mathrm{d}x^6.
\end{eqnarray}
Now, for the orbifolding to preserve the metric some of the $R^A$ must
be set equal to one another. It is straightforward to check that if
$\alpha_\tau$ involves a rotation in the $(A,B)$ plane by an angle not equal
to $\pi$, then we must set $R^A=R^B$. Following this prescription it
is easy to find the function $b(\Gamma)$ discussed above.

On one of the blow-ups $T^3\times U$ (for convenience we suppress
$\tau$ and $n$ indices) we use coordinates $\xi^a$ on $T^3$ and
four-dimensional coordinates $\zeta^\mu$ on $U$. We write $R^a$ to
denote the three radii of $T^3$, which will be the three $R^A$ in the
directions fixed by $\alpha$. The $G_2$ structure can be written as
\begin{equation} \label{structure4}
\varphi=\sum_a\omega^a\left(w(\zeta),z(\zeta),\boldsymbol{b}_1,\ldots,\boldsymbol{b}_{N_\tau}\right)\wedge R^a\mathrm{d}\xi^a-R^1R^2R^3\mathrm{d}\xi^1\wedge\mathrm{d}\xi^2\wedge\mathrm{d}\xi^3.
\end{equation}
Here $w$ and $z$ are complex coordinates on $U$ and the
$\boldsymbol{b}_i\equiv(\mathrm{Re}\:a_i,\mathrm{Im}\:a_i,b_i)$ are a
set of three-vectors, which parameterize the size of the blow-up and
its orientation with respect to the bulk. The $\omega^a$ are a triplet
of two-forms that constitute a ``nearly'' hyperk\"ahler structure on
$U$, as discussed in Appendix \ref{c}. We will not need to know
explicitly the relation between the two sets of coordinates
$\zeta^\mu$ and $w$ and $z$, although we keep in mind that this
relation will depend on the four radii $R^\mu$ transverse to the
$R^a$. In terms of $w$ and $z$, we can write the $\omega^a$ as
\begin{equation} \label{om1}
\omega^1=\frac{i}{2}\partial\bar{\partial}\mathcal{K},
\end{equation}
\begin{equation} \label{om2}
\omega^2=-\mathrm{Re}\left(\frac{\mathrm{d}w\wedge\mathrm{d}z}{w}\right), \: \: \:
\omega^3=-\mathrm{Im}\left(\frac{\mathrm{d}w\wedge\mathrm{d}z}{w}\right),
\end{equation}
where $\mathcal{K}$ is the K\"ahler potential for $U$, which
interpolates between that for Gibbons-Hawking space in the central
region of the blow-up and that for flat space far away from the centre
of the blow-up. We clarify this statement in the discussion below, but
first we shall describe the central region of the blow-up, where $U$
looks exactly like Gibbons-Hawking space. For a technical account of
this discussion, including how to write $\mathcal{K}$ explicitly, we
refer the reader to Appendix \ref{c}.

Gibbons-Hawking spaces (or gravitational multi-instantons) provide a
generalization of the Eguchi-Hanson space and their different
topological types are labelled by an integer $N$ (where the case $N=2$
corresponds to the Eguchi-Hanson case). While the Eguchi-Hanson space
contains a single two-cycle, the $N^{\rm th}$ Gibbons-Hawking space
contains a sequence $\gamma_1,\ldots,\gamma_{N-1}$ of such cycles at
the ``centre'' of the space.  Only neighbouring cycles $\gamma_i$ and
$\gamma_{i+1}$ intersect and in a single point and, hence, the
intersection matrix $\gamma_i\cdot\gamma_j$ equals the Cartan matrix of
$A_{N-1}$. Asymptotically, the $N^{\rm th}$ Gibbons-Hawking space has
the structure $\mathbb{C}^2/\mathbb{Z}_{N}$.  Accordingly, we take
$N=N_\tau$ when blowing up $\mathbb{C}^2/\mathbb{Z}_{N_{\tau}}$. The
metric on Gibbons-Hawking space can be written
\begin{equation}
\mathrm{d}s^2=\gamma\mathrm{d}z\mathrm{d}\bar{z} + \gamma^{-1} \left( \frac{\mathrm{d}w}{w}+\bar{\delta}\mathrm{d}z \right) \left( \frac{\mathrm{d}w}{w}+ \delta\mathrm{d}\bar{z} \right),
\end{equation}
where
\begin{equation}
\gamma = \sum_i\frac{1}{r_i},
\end{equation}
\begin{equation}
r_i=\sqrt{(x-b_i)^2+4\lvert z-a_i\rvert^2},
\end{equation}
\begin{equation}
\delta =  \sum_i\frac{x-b_i-r_i}{2(\bar{z}-\bar{a}_i)r_i},
\end{equation}
\begin{equation}
w\bar{w}=\prod_i(x-b_i+r_i).
\end{equation}
Here $x$ is a real coordinate, given implicitly in terms of $w$ and
$z$ in the above equations. The sizes and orientations of the
two-cycles are determined by the $N_\tau$ points $\boldsymbol{b}_i$ in
the $(\mathrm{Re}\:z,\mathrm{Im}\:z,x)$ hyperplane. Concretely
$\gamma_i$ is parameterized by
\begin{equation}
z=a_i+\lambda (a_{i+1}-a_i),
\end{equation}
\begin{equation}
w=e^{i\theta}h(\lambda),
\end{equation} 
for some function $h$, as $0\leq\lambda\leq 1$, $0\leq\theta\leq 2\pi$ \cite{Hitchin}.

We can add a periodic real coordinate $y$ to $x$, $z$ and $\bar{z}$ to form a well-defined coordinate system on the space. We can also define a radial coordinate $r$ given by
\begin{equation}
r= \sqrt{(x-\tilde{b})^2+4\lvert z-\tilde{a} \rvert^2},
\end{equation}
where tildes denote mean values over the index $i$. It is precisely
this radial coordinate that the interpolating function $\epsilon$,
appearing in the K\"ahler potential $\mathcal{K}$, depends on. We can
now describe the interpolation more precisely. If we set all the $a$
and $b$ parameters of Gibbons-Hawking space to zero we have flat
space. Therefore the method of constructing $\mathcal{K}$ is to start
with the K\"ahler potential for Gibbons-Hawking space, and to then
place a factor of $\epsilon$ next to every $a$ and $b$ that
appears. We can keep $\epsilon$ general, all we require are the
following properties:
\begin{equation}
\epsilon (r) = \left\{ \begin{array}{cc}
1 & \mathrm{if} \; r\leq r_0,  \\
0 & \mathrm{if} \; r\geq r_1, \\
\end{array} \right.
\end{equation}
where $r_0 $ and $r_1$ are two fixed radii satisfying $\lvert
a_i\rvert\ll r_0 < r_1$ and $\lvert b_i\rvert\ll r_0$ for each
$i$. Then, as already discussed, $U$ is identical to Gibbons-Hawking
space for $r<r_0$. For $r>r_1$ $U$ is identical to the flat space
$\mathbb{C}^2/\mathbb{Z}_N$, and we can match the $G_2$ structure
\eqref{structure4} to the bulk $G_2$ structure \eqref{structure3}.

Let us briefly discuss the torsion of the $G_2$ structure
\eqref{structure4}. By virtue of the blow-up $U$ being hyperk\"ahler
in the regions $r<r_0$ and $r>r_1$, and the $\omega^a$ of equations
\eqref{om1} and \eqref{om2} forming the triplet of closed and
co-closed K\"ahler forms expected on such a space, the $G_2$ structure
is torsion free in these regions. It departs from non-zero torsion
only in the ``collar'' region $r\in[r_0,r_1]$, where $\omega^2$ and
$\omega^3$ fail to be co-closed. However, for sufficiently small
blow-ups, $\lvert a_i\rvert\ll 1$, $\lvert b_i\rvert\ll 1$, and a
``smooth'', slowly-varying interpolation function $\epsilon$, the
deviation from a torsion-free $G_2$ structure is small
\cite{Lukas}. Consequently, we can use this $G_2$ structure to
reliably compute the K\"ahler potential to leading non-trivial order
in the $a_i$s and $b_i$s.

We end this section by briefly discussing the metric on a blow-up. The
metric can be derived directly from the $G_2$ structure, using
equations \eqref{met1} and \eqref{met2}. Its structure is given by
\begin{equation}
\mathrm{d}s^2=\mathcal{G}_0\mathrm{d}\boldsymbol{\zeta}^2+\sum_{a=1}^{3}\mathcal{G}_a(\mathrm{d}\xi^a)^2,
\end{equation}
where $\mathrm{d}\boldsymbol{\zeta}$ is the line element on the
appropriate smoothed Gibbons-Hawking space, which can be derived from
equations \eqref{Legendre} and \eqref{smoothF}, and the $\mathcal{G}$s
are conformal factors, that may depend on the blow-up moduli and the
interpolation function, but whose product must be equal to 1, since
they do not appear in the measure \eqref{measure}.

\section{Periods, Volumes and K\"ahler Potentials} \label{kaehler}
Having written down a $G_2$ structure of small torsion on our general
manifold $\mathcal{M}$, we can now compute the periods. The bulk
periods
\begin{equation}
a^A=\int_{C^A}\varphi
\end{equation}
are straightforward to obtain and are given by
\begin{equation} \label{period2}
\left. \begin{array}{cccc}
a^1=R^1R^2R^3, & a^2=R^1R^4R^5, & a^3=R^1R^6R^7, & a^4=R^2R^4R^6, \\
a^5=R^2R^5R^7, & a^6=R^3R^4R^7, & a^7=R^3R^5R^6. &  \, \\
\end{array} \right.
\end{equation}
To find the periods associated with the blow-ups, we firstly require
the period integrals
\begin{equation}
\int_{\gamma_i}\omega^a,
\end{equation}
$i=1,\ldots,N-1$ on a general Gibbons-Hawking space. Following
Ref. \cite{Hitchin} we find
\begin{equation}
\int_{\gamma_i}\omega^1=\frac{\pi}{2}(b_i-b_{i+1}),
\end{equation}
\begin{equation}
\int_{\gamma_i}(\omega^2+i\omega^3)=\pi i(a_i-a_{i+1}).
\end{equation}
Having obtained these we can write down the periods
\begin{equation}
A(\tau,n,a,i)=\int_{C(\tau,n,a,i)}\varphi
\end{equation}
on each blow-up $U_{(\tau,n)}\times T^3_{(\tau,n)}$. We find
\begin{eqnarray} \label{period1}
A(\tau,n,1,i) & = & \frac{\pi}{2} R^1_{(\tau)}\left( b_{(\tau,n,i)} - b_{(\tau,n,i+1)} \right), \nonumber \\
A(\tau,n,2,i) & = & \frac{i\pi}{2}  R^2_{(\tau)}\left( a_{(\tau,n,i)}           - \bar{a}_{(\tau,n,i)}      - a_{(\tau,n,i+1)}  + \bar{a}_{(\tau,n,i+1)}  \right), \nonumber \\
A(\tau,n,3,i) & = & \frac{\pi}{2} R^3_{(\tau)}\left( a_{(\tau,n,i)}           + \bar{a}_{(\tau,n,i)}      - a_{(\tau,n,i+1)}  - \bar{a}_{(\tau,n,i+1)}  \right).
\end{eqnarray}
We remind the reader that $R^a_{(\tau)}$ denote the three radii of
$T^3_{(\tau,n)}$, consistent with the notation of equation
\eqref{structure4}.  \\

Our next task is to find the total volume of $\mathcal{M}$. From the
bulk metric \eqref{bulkmetric}, we see that the bulk contribution to
the volume is proportional to $\prod_A R^A$. There will be a factor
$f(\Gamma)$ in front of this, dependent on the orbifold group
$\Gamma$. In certain simple cases this is just the inverse of the
order of $\Gamma$, and it is always calculable by obtaining the
fundamental domain of the orbifold. For the purposes of most
calculations, it is relatively unimportant, since one will be able to
absorb it into the normalization of the blow-up moduli fields. The
contributions to the volume from the blow-ups are easily obtainable
from equations \eqref{measure} and \eqref{finalvol}. Putting
everything together we find, to lowest non-trivial order in the blow
up moduli,
\begin{equation} \label{volM}
V =   f(\Gamma)\prod_AR^A  -   \frac{\pi^2}{6}\sum_{\tau,n}  N_\tau\big(  \mathrm{var}_i\{b_{(\tau,n,i)}\}+ 2\,\mathrm{var}_i\{\mathrm{Re}\,a_{(\tau,n,i)}\} + 2\, \mathrm{var}_i\{ \mathrm{Im}\,a_{(\tau,n,i)}\}\big)\prod_a R^a_{(\tau)}.
\end{equation}
Here var refers to the variance, with the usual definition:
\begin{equation}
\mathrm{var}_i\{X_i\}=\frac{1}{N}\sum_i(X_i-\tilde{X})^2.
\end{equation}
Note that this result is independent of the interpolation functions
$\epsilon_{(\tau,n)}$.  \\
 
We are now ready to compute the K\"ahler potential. Using the results
\eqref{period1} and \eqref{period2} for the periods, we can rewrite
the volume \eqref{volM} in terms of $a^A$ and $A(\tau,n,a,i)$, which
constitute the real, bosonic parts of superfields. We denote these
superfields by $T^A$ and $U^{(\tau,n,a,i)}$ such that
\begin{equation}
\mathrm{Re}(T^A)=a^A, \; \; \; \mathrm{Re}(U^{(\tau,n,a,i)})=A(\tau,n,a,i).
\end{equation}
Note that in many cases some of the $T^A$s are identical to each other
and should be thought of as the same field. As discussed in Section
\ref{manifolds}, this comes about from requiring some of the radii
$R^A$ to be equal for a consistent orbifolding of the base torus. The
number of distinct $T^A$ is $b(\Gamma)$, which for the orbifold groups
$\Gamma$ constructed in Section \ref{groups} is given in Table~1. To
discover which $T^A$ are equal the procedure is as follows. Let $x^A$
be coordinates in the bulk with respect to which the $G_2$ structure
is given by \eqref{structure3}.  Then a generator $\alpha_\tau$ of
$\Gamma$ acts by simultaneous rotations in two planes $(A,B)$ and
$(C,D)$ say. If the order of $\alpha_\tau$ is greater than two, then
we identify $R^A$ with $R^B$ and $R^C$ with $R^D$.  We go through this
process for all generators of $\Gamma$ and then use \eqref{period2} to
determine which of the $a^A$ and hence $T^A$ are equal.  From
Eq.~\eqref{witten} we find for the K\"ahler potential
\begin{equation} \label{K}
K =  -\sum_{A=1}^{7}\ln (T^A+\bar{T}^A) -3\ln \left[ 1-\frac{2}{3f(\Gamma)}\sum_{n,\tau,a}\frac{1}{N_\tau}\frac{\sum_{i<j}\left(\sum_{k=i}^{j-1}(U^{(\tau,n,a,k)}+\bar{U}^{(\tau,n,a,k)})\right)^2}{(T^{A(\tau,a)}+\bar{T}^{A(\tau,a)})(T^{B(\tau,a)}+\bar{T}^{B(\tau,a)})}\right] + c,
\end{equation}
where the constant c is given by
\begin{equation}
c=10\ln 2-3\ln f(\Gamma) + 6\ln\pi.
\end{equation}
The index functions $A(\tau,a)$, $B(\tau,a)\in\{1,\ldots,7\}$ indicate
by which two of the seven bulk moduli $T^A$ the blow up moduli
$U^{(\tau,n,a,i)}$ are divided in the K\"ahler potential
\eqref{K}. Their values depend only on the generator index $\tau$ and the
orientation index $a$. They may be calculated from the formula
\begin{equation}
a^{A(\tau,a)}a^{B(\tau,a)}=\frac{\left(R^a_{(\tau)}\right)^2\prod_AR^A}{\prod_bR^b_{(\tau)}}.
\end{equation}
The $\tau$ dependence is only through the fixed directions of the generator $\alpha_\tau$ and the possible values of the index functions are given in Table~2. 
\begin{table}
\begin{center}
\begin{tabular}{|c|c|c|c|}
\hline
$\mathrm{Fixed \, directions\, of\, } \alpha_\tau$ &$ a=1$ & $a=2$ &$ a=3$ \\
\hline
(1,2,3) & (2,3) & (4,5) & (6,7) \\
\hline
(1,4,5) & (1,3) & (4,6) & (5,7) \\
\hline
(1,6,7) & (1,2) & (4,7) & (5,6) \\
\hline
(2,4,6) & (1,5) & (2,6) & (3,7) \\
\hline
(2,5,7) & (1,4) & (2,7) & (3,6) \\
\hline
(3,4,7) & (1,7) & (2,4) & (3,5) \\
\hline
(3,5,6) & (1,6) & (2,5) & (3,4) \\
\hline
\end{tabular}
\caption{Values of the index functions $(A(\tau,a),B(\tau,a))$ specifying the bulk moduli $T^A$ by which the blow-up moduli $U^{(\tau,n,a,i)}$ are divided in the K\"ahler potential.}
\end{center}
\end{table}

We now state precisely and systematically the scenarios in which
\eqref{K} is valid. Firstly all moduli must be larger than one (in
units where the Planck length is set to one) so that the supergravity
approximation to M-theory is valid. Secondly, all blow-up moduli
$U^{(\tau,n,a,i)}$ must be small compared to the bulk moduli $T^A$ so
that corrections of higher order in $U/T$ can be neglected. The action
of the generators of the orbifold group $\Gamma$ on the base
seven-torus $T^7$ must lead to an orbifold $\mathcal{O}$ with
co-dimension four singularities. Furthermore, no two generators must
fix the same point on the torus. This last requirement is to ensure
that the assumption of the structure
$T^3\times\mathbb{C}^2/\mathbb{Z}_N$ around the singularities is
correct.


\section{Conclusion}
Let us summarize what we have found, and mention some possible
extensions. Equation \eqref{K} gives the moduli K\"ahler potential for
a large class of compact manifolds of holonomy $G_2$. This class
contains manifolds constructed from a large number of different
orbifolds, based on at least sixteen distinct orbifold groups, namely
those constructed in Section \ref{groups}. Moduli fields fall into two
categories. Firstly the fields $T^A$, which descend from the seven
radii of the manifold, and secondly the fields $U^{(\tau,n,a,i)}$
which descend from geometrical parameters describing the blow-up of
singularities of the manifold (namely the radii of two-cycles on the
blow-up and their orientation with respect to the bulk). Our formula
constitutes the first two terms in an expansion of the K\"ahler
potential in terms of the $U$s. The zeroth order term is simply a
consequence of the volume of the manifold being proportional to the
product of the seven radii. It is not surprising that the lowest order
correction terms arise at second order in the $U$s. Heuristically, one
can think of all $U$ dependent terms as being associated with the
volume subtracted from the manifold as a result of the presence of
two-cycles on the blow-ups. One expects these terms to depend on the
two-cycles through their area, and hence on even powers of the
$U$s. Dimensional analysis insists that all second-order terms in the
$U$s are homogeneous of order minus two in the $T$s, but it did not
have to necessarily turn out that the terms took on so simple a form
in the general case. This is an attractive feature of our result.

An interesting extension of the work in this paper would be to attempt
to include conical, or co-dimension seven, singularities onto the
manifolds, thus supporting charged chiral matter. We found a large number
of orbifold group elements that lead to co-dimension six
singularities. One could attempt to generalize the blow-up procedure
and K\"ahler potential calculation to the case of manifolds with
orbifold groups containing such elements. There is also the
possibility of a more complicated orbifold fixed point structure. For
example, if there exist points on the torus that are fixed by more
than one generator of the orbifold group, there can be several
topologically distinct ways of blowing up the associated singularity
\cite{Joyce}. Questions of moduli stabilization can also now be looked
at in a more general context by application of our K\"ahler potential
formula.

\section*{Acknowledgements}
We would like to thank James Gray for helpful discussion. A.~B.~B.~is
supported by a PPARC Postgraduate Studentship. A.~L.~is supported by a
PPARC Advanced Fellowship.

\section*{Appendix}

\appendix

\section{Some results about $G_2$ structures} \label{a}
In this appendix we derive some results about $G_2$ structures that
will be used in our calculations.  \\

Let $A$ be an element of $SO(7)$ of the form
\begin{equation} \label{a1}
A=\left( \begin{array}{cccc}
1 & \, & \, & \,  \\
\, & R(\theta_{1}) & \, & \,  \\
\, & \, & R(\theta_{2}) & \, \\
\, & \, & \, & R(\theta_{3})  \\
\end{array} \right) ,  
\end{equation} 
where
\begin{equation}
 R(\theta_{i})= \left( \begin{array}{cc}
\cos\theta_{i} & -\sin\theta_{i} \\
\sin\theta_{i} & \cos\theta_{i} \\
\end{array} \right).
\end{equation}
Then $A$ is in $G_2$ (for some embedding of $G_2$ into $SO(7)$) if and
only if one of the following conditions hold on the $\theta_i$:
\begin{eqnarray}
 \theta_{1}+\theta_{2}+\theta_{3} & = & 0 \: \mathrm{mod} \: 2\pi, \label{1}\\
  -\theta_{1}+\theta_{2}+\theta_{3} & = & 0 \: \mathrm{mod} \: 2\pi, \label{2}\\
  \theta_{1}-\theta_{2}+\theta_{3} & = & 0 \: \mathrm{mod} \: 2\pi, \label{3}\\
  \theta_{1}+\theta_{2}-\theta_{3} & = & 0 \: \mathrm{mod} \: 2\pi. \label{4}
\end{eqnarray}

\textit{Proof:} That this is sufficient has already been demonstrated
in Section \ref{elements}. Now let us assume that this is not
necessary and we will find a contradiction. We shall attempt to
construct a three-form $\varphi$ that defines a $G_2$ structure and
that is left invariant by some $A$ not satisfying one of \eqref{1},
\eqref{2}, \eqref{3} or \eqref{4}.

When expressed in terms of the coordinates $x_{0},$ $z_{1},$ $z_{2}$
and $z_{3}$, as in \eqref{complex}, each non-vanishing component of a
three-form imposes a definite constraint on the angles $\theta_{i}$ if
it is to be left invariant by $A$. For example a three-form with a
non-vanishing coefficient of
$\mathrm{d}x_{0}\wedge\mathrm{d}z_{1}\wedge\mathrm{d}z_{2}$ imposes
the constraint $\theta_1+\theta_2=0$. We shall use this property
whilst attempting to construct our $G_{2}$ structure.

Since the $G_{2}$ structure $\varphi$ given in \eqref{structure}
satisfies the following tensorial identity, so must the $G_{2}$
structure that we are attempting to construct.
\begin{equation}
\varphi_{mnr}\varphi_{pq}^{\phantom{pq}r}=\phi_{mnpq}+\delta_{mp}\delta_{nq}-\delta_{mq}\delta_{np},
\end{equation}
where $\phi_{mnpq}$ is the four-form dual to $\varphi$, and
$m,n,\ldots=1,\ldots,7$ label the seven real dimensions of the
manifold. From this equation we obtain
\begin{equation} \label{constraint1}
\sum_{p=1}^{7}\varphi_{pqr}^{2}=1,
\end{equation}
for $q,r=1,2,\ldots,7$. Now in order for invariance of $\varphi$ to
not impose any of \eqref{one}, \eqref{two}, \eqref{three} or
\eqref{four} on $A$, when expressed in the coordinates of
\eqref{complex} it must not contain any of the following terms
\begin{equation}
\mathrm{d}z_{1}\wedge\mathrm{d}z_{2}\wedge\mathrm{d}z_{3}, \: \: \mathrm{d}\bar{z}_{1}\wedge\mathrm{d}z_{2}\wedge\mathrm{d}z_{3}, \: \: \mathrm{d}z_{1}\wedge\mathrm{d}\bar{z}_{2}\wedge\mathrm{d}z_{3}, \: \: \mathrm{d}z_{1}\wedge\mathrm{d}z_{2}\wedge\mathrm{d}\bar{z}_{3}.
\end{equation}
For this to be the case, all of the following must be zero:
\begin{equation}
\varphi_{246}, \: \varphi_{247}, \: \varphi_{256}, \: \varphi_{257}, \: \varphi_{346}, \: \varphi_{347}, \: \varphi_{356}, \: \varphi_{357}.
\end{equation}
Bearing this in mind, and taking $(q,r)=(4,6)$ in equation
\eqref{constraint1} we observe that at least one of the following must
be non-zero:
\begin{equation}
\varphi_{146}, \: \varphi_{546}, \: \varphi_{746}, \: 
\end{equation}
By changing coordinates to those of \eqref{complex}, we can spot the
constraints this imposes on the $\theta_{i}$. We can repeat this
process for $(q,r)=(2,4)$ and $(q,r)=(2,7)$, and put all the
constraints together to obtain the result that $A$ only leaves
$\varphi$ invariant if at least two of the $\theta_{i}$ are zero. By
assumption $\varphi$ is left invariant by some $A$ not satisfying any
of \eqref{one}, \eqref{two}, \eqref{three} or \eqref{four}, and so
assume, by the above, and without loss of generality, that this $A$
has $\theta_{1}=\theta_{2}=0,$ and $\theta_{3}\neq0$ mod $2\pi$. Then
it is easily verified that only the following components (and
components with permuted indices of those below) of $\varphi$ may be
non-zero:
\begin{equation}
\begin{array}{l}
\varphi_{123}, \: \varphi_{145}, \: \varphi_{167}, \: \varphi_{124}, \: \varphi_{125}, \:\varphi_{134}, \: \varphi_{135}, \: \varphi_{234}, \\  \varphi_{235}, \: \varphi_{245},  \: \varphi_{345}, \: \varphi_{267}, \: \varphi_{367}, \: \varphi_{467}, \: \varphi_{567}.
\end{array}
\end{equation}
Now, using \eqref{constraint1} we find that
\begin{equation} \label{constraint2}
\lvert\varphi_{467}\rvert=\lvert\varphi_{567}\rvert=\lvert\varphi_{267}\rvert=\lvert\varphi_{367}\rvert=1,
\end{equation}
by using for example $(q,r)=(4,6)$. We then invoke the identity
\begin{equation}
\varphi_{mpq}\varphi^{npq}=6\delta_{m}^{n},
\end{equation}
which implies
\begin{equation}
\varphi_{7pq}\varphi^{7pq}=6.
\end{equation}
However \eqref{constraint2} gives 
\begin{eqnarray}
\varphi_{7pq}\varphi^{7pq} & \geq & 2(\varphi_{746}\varphi^{746}+\varphi_{756}\varphi^{756}+\varphi_{726}\varphi^{726}+\varphi_{736}\varphi^{736}) \nonumber\\
&= & 8.
\end{eqnarray}
We therefore have a contradiction, and hence our result.
\\

Now let $A$ be as in \eqref{a1}, and let it satisfy one of the conditions \eqref{1}-\eqref{4} so that it preserves some $G_2$ structure, but now let us assume that $\theta_1=0$ for simplicity. Then any $G_2$ structure preserved by $A$ may be brought to the standard form of equation \eqref{structure}, up to possible sign differences, by some redefinition of coordinates that preserves the structure of $A$ up to some redefinition of $\theta_2$ and $\theta_3$.

\textit{Proof:} Following the method used to prove the previous result, we can draw up a list of the components of $\varphi$ that may be non-zero if it is to be preserved by $A$. These are 
\begin{equation}
\begin{array}{l}
\varphi_{123}, \: \varphi_{145}, \: \varphi_{167}, \: \varphi_{146}, \: \varphi_{147}, \:\varphi_{156}, \: \varphi_{157}, \: \varphi_{245}, \: \varphi_{345}, \: \varphi_{267}, \\  \varphi_{367}, \: \varphi_{246}, \: \varphi_{247}, \: \varphi_{256}, \: \varphi_{257}, \: \varphi_{346}, \: \varphi_{347},\: \varphi_{356}, \: \varphi_{357}.
\end{array}
\end{equation}
Now using \eqref{constraint1} and our freedom to choose the orientation of the 1, 2 and 3 directions we see that we have $\varphi_{123}=1$. We can then show similarly that, without loss, $\varphi_{145}=1$. Then $\varphi_{167}=\pm 1$ by consistency of the following identity, with $(m, n, q, r, s)=(2,6,2,3,6)$:
\begin{equation}
\varphi_{mn}^{\phantom{mn}p}\phi_{pqrs}=\delta_{mq}\varphi_{nrs}+\delta_{mr}\varphi_{nsq}+\delta_{ms}\varphi_{nqr}-\delta_{nq}\varphi_{mrs}-\delta_{nr}\varphi_{msq}-\delta_{ns}\varphi_{mqr}.
\end{equation}
Finally, repeated use of \eqref{constraint1} and remaining coordinate freedom enables us to establish the result.
\\

There is a useful corollary of the above result: Let $A$ be a rotation matrix with three independent preserved directions $p$, $q$ and $r$. Then $A$ preserves a given $G_2$ structure $\varphi$ only if $\lvert\varphi_{pqr}\rvert=1$.

\section{Table of Possible Orbifold Group Elements of $G_{2}$ Manifolds} \label{b}
The possible generators $\alpha$ of orbifold groups take the form $\alpha : \boldsymbol{x} \mapsto A_{(\alpha)}\boldsymbol{x} + \boldsymbol{b}_{(\alpha)}$, where $A_{(\alpha)}$ is an orthogonal matrix, which can be put into block diagonal form 
\begin{equation}
\left( \begin{array}{cccc}
1 & \, & \, & \,  \\
\, & R(\theta_{1}) & \, & \,  \\
\, & \, & R(\theta_{2}) & \, \\
\, & \, & \, & R(\theta_{3})  \\
\end{array} \right) ,  
\end{equation} 
where
\begin{equation}
 R(\theta_{i})= \left( \begin{array}{cc}
\cos\theta_{i} & -\sin\theta_{i} \\
\sin\theta_{i} & \cos\theta_{i} \\
\end{array} \right),
\end{equation}
and the possibilities for the $\theta_i$ (up to signs) are listed in Table~3. The $n_i$ and the $d_i$ label how the representation $R$ defined by $A_{(\alpha)}$ decomposes into irreducibles according to $R=n_1R_1\oplus\cdots\oplus n_rR_r$, as in Section \ref{groups}.
\begin{center}
\begin{tabular}{c||c||cc|cc|cc|cc}
$\mathrm{Symmetry}$ & $\frac{1}{2\pi}(\theta_{1},\theta_{2},\theta_{3})$  & $n_{1}$ & $d_{1}$ & $n_{2}$ & $d_{2}$ & $n_{3}$ & $d_{3}$ & $n_{4}$ & $d_{4}$  \\ \hline
$\mathbb{Z}_2$& $(0,\frac{1}{2},\frac{1}{2})$  & 3 & 1 & 4 &1&-&-&-&- \\
$\mathbb{Z}_3$ &$ (0,\frac{1}{3},\frac{1}{3})$  & 3 & 1 & 2 &2&-&-&-&- \\
$\mathbb{Z}_3^\ast$ & $(\frac{1}{3},\frac{1}{3},\frac{1}{3})$ & 1 & 1 & 3 &2&-&-&-&- \\
$\mathbb{Z}_4$ & $(0,\frac{1}{4},\frac{1}{4})$  & 3 & 1 & 2 &2&-&-&-&- \\
$\mathbb{Z}_4^\ast$ &$ (\frac{1}{4},\frac{1}{4},\frac{1}{2})$  & 1 & 1 & 2 &2&2&1&-&- \\
$\mathbb{Z}_6$ &$ (0,\frac{1}{6},\frac{1}{6})$  & 3 & 1 & 2 &2&-&-&-&- \\
$\mathbb{Z}_6^\ast$ &$ (\frac{1}{6},\frac{1}{6},\frac{1}{3}) $ & 1 & 1 & 2 &2&1&2&-&- \\
$\mathbb{Z}_6^\dagger$ &$ (\frac{1}{6},\frac{1}{3},\frac{1}{2}) $ & 1 & 1 & 1 & 2 & 1& 2 & 2 & 1 \\
$\mathbb{Z}_7$ &$ (\frac{1}{7},\frac{2}{7},\frac{3}{7})$  & 1 & 1 & 1 & 2 & 1& 2 & 1 & 2 \\
$\mathbb{Z}_8$ &$ (\frac{1}{8},\frac{1}{4},\frac{3}{8}) $ & 1 & 1 & 1 & 2 & 1& 2 & 1 & 2 \\
$\mathbb{Z}_8^\ast$ &$ (\frac{1}{8},\frac{3}{8},\frac{1}{2})$  & 1 & 1 & 1 & 2 & 1& 2 & 2 & 1 \\
$\mathbb{Z}_{12}$ & $(\frac{1}{12},\frac{1}{3},\frac{5}{12}) $ & 1 & 1 & 1 & 2 & 1& 2 & 1 & 2 \\
$\mathbb{Z}_{12}^\ast$ & $(\frac{1}{12},\frac{5}{12},\frac{1}{2}) $ & 1 & 1 & 1 & 2 & 1& 2 & 2 & 1 \\
\end{tabular}
\end{center}

\section{Blow-up and some Calculations on Gibbons-Hawking Space} \label{c}
In this Appendix we begin by deriving a general volume formula valid
on regions of $G_2$ manifolds that take the form $T^3\times U$, where
$T^3$ is some three-torus and $U$ is a four dimensional hyperk\"ahler
space. We then describe, for arbitrary $N$, how to blow-up
$T^3\times\mathbb{C}^2/\mathbb{Z}_N$, and use our formula to compute
volumes on blow-ups, as induced by a given $G_2$ structure. We
consider in turn cases in which $U$ approaches flat space
asymptotically and in which $U$ becomes exactly flat for sufficiently
large radius. We follow and generalize results from Ref. \cite{Lukas}.
\\

Let us briefly recall the definition of a hyperk\"ahler space. A hyperk\"ahler space is a $4m$-dimensional Riemmanian manifold admitting a triplet $J^a$ of covariantly constant complex structures satisfying the algebra
\begin{equation}
J^aJ^b=-\boldsymbol{1}\delta^{ab}+\epsilon^{ab}_{\phantom{ab}c}J^c.
\end{equation}
Associated with the $J^a$ via \suppressfloats[t]
\begin{equation}
\omega^a_{\phantom{a}\mu\nu}=(J^a)_\mu^{\phantom{\mu}\rho}g_{\rho\nu}
\end{equation}
we have a triplet $\omega^a$ of covariantly constant so-called K\"ahler forms.

If we let $U$ be a hyperk\"ahler space, then we can write down the following $G_2$ structure on $T^3\times U$:
\begin{equation} \label{struc2}
\varphi=\sum_a\omega^a\wedge\mathrm{d}\xi^a-\mathrm{d}\xi^1\wedge\mathrm{d}\xi^2\wedge\mathrm{d}\xi^3,
\end{equation}
where $\xi^a$ are coordinates on the torus $T^3$. This is torsion free by virtue of $\mathrm{d}\omega^a=\mathrm{d}\star\omega^a=0$.

The volume element $\sqrt{\mathrm{det}(g)}$ on $T^3\times U$ may be found from the $G_2$ structure via the equations
\begin{equation} \label{met1}
g_{AB}=\mathrm{det}(\gamma)^{-1/9}\gamma_{AB}, \: \: \: \sqrt{\mathrm{det}(g)}=\mathrm{det}(\gamma)^{1/9},
\end{equation}
where
\begin{equation} \label{met2}
\gamma_{AB}=\frac{1}{144}\varphi_{ACD}\varphi_{BEF}\varphi_{GHI}\hat{\epsilon}^{CDEFGHI},
\end{equation}
where $\hat{\epsilon}$ is the ``pure-number'' Levi-Civita pseudo-tensor.

We now follow a general method of construction of hyperk\"ahler spaces \cite{Hitch1} to derive a formula for the measure. The triplet of K\"ahler forms is given by
\begin{equation}
\omega^1=\frac{i}{2}\partial\bar{\partial}\mathcal{K},
\end{equation}
\begin{equation}
\omega^2=\mathrm{Re}(\mathrm{d}u\wedge\mathrm{d}z), \: \: \: 
\omega^3=\mathrm{Im}(\mathrm{d}u\wedge\mathrm{d}z),
\end{equation}
where $\mathcal{K}$ is the K\"ahler potential for $U$. These give
\begin{equation} \label{det}
\mathrm{det}\gamma=\frac{1}{4^9}(\mathcal{K}_{,u\bar{z}}\mathcal{K}_{,z\bar{u}}-\mathcal{K}_{,u\bar{u}}\mathcal{K}_{,z\bar{z}})^3.
\end{equation}
Here $u$ and $z$ are complex coordinates on $U$ and $\mathcal{K}_{,u\bar{z}}\equiv\partial^2\mathcal{K}/\partial u\partial\bar{z}$ etc. We can reduce this to a simpler expression if we write $\mathcal{K}$ as the Legendre transform of a real function $\mathcal{F}(x,z,\bar{z})$ with respect to the real coordinate $x$:
\begin{equation} \label{Legendre}
\mathcal{K}(u,\bar{u},z,\bar{z})=\mathcal{F}(x,z,\bar{z})-(u+\bar{u})x.
\end{equation}
Here $x$ is a function of $z$, $\bar{z}$, $u$ and $\bar{u}$ determined by
\begin{equation} \label{u}
\frac{\partial \mathcal{F}}{\partial x}=u+\bar{u}.
\end{equation}
We can then re-express \eqref{det} in terms of partial derivatives of $\mathcal{F}$ and obtain the rather neat result that
\begin{equation} \label{measure}
\sqrt{\mathrm{det}(g)}\equiv \frac{1}{4}.
\end{equation}
Note that this is entirely general and valid on any region of a $G_2$
manifold on which the $G_2$ structure can be written as in
Eq.~\eqref{struc2}, and on which the K\"ahler potential can be
expressed as in \eqref{Legendre}. Eq.~\eqref{measure} gives the measure
for integrating over the coordinates $u$, $z$, $\bar{u}$, $\bar{z}$
and $\xi^a$. However it will be more convenient in what follows for us
to substitute $u$ and $\bar{u}$ for the real coordinates $x$ and $y$,
with $x$ as in \eqref{u} and $y$ given by $y=i(\bar{u}-u)$. Then,
assuming for convenience unit volume for $T^3$, the volume of
$T^3\times U$ over some compact subspace $U_0\subset U$ is given by
\begin{equation} \label{integral}
\mathrm{vol}(U_0)=\frac{1}{8}\int_{U_0}\lvert\mathcal{F}_{,xx}\mathrm{d}z\mathrm{d}\bar{z}\mathrm{d}x\mathrm{d}y\rvert.
\end{equation}
\\

As suggested by the above, the blow-up of
$T^3\times\mathbb{C}^2/\mathbb{Z}_N$ will take the form $T^3\times U$,
where $U$ is an appropriate hyperk\"ahler space. More specifically,
$U$ will belong to the family of spaces referred to as Gibbons-Hawking
spaces or ``gravitational multi-instantons'' \cite{Hitchin}. Note that
these are generalised versions of Eguchi-Hanson space, which
corresponds to the case of $N=2$. For each $N$, we may take $U$ to be
the $N$-centred Gibbons-Hawking space, for which the function
$\mathcal{F}$ is given by
\begin{equation} \label{potential}
\mathcal{F}=\sum_{i=1}^{N}\bigg(r_i-x_i\mathrm{ln}(x_i+r_i)+\frac{x_i}{2}\ln(4z_i\bar{z}_i)\bigg),
\end{equation}
where
\begin{equation}
x_i=x-b_i, \: \: z_i=z-a_i,
\end{equation}
\begin{equation}
r_i=\sqrt{x_i^2+4\lvert z_i\rvert^2}.
\end{equation}
We can derive the metric from $\mathcal{F}$ by using the expression \eqref{Legendre} for the K\"ahler potential. In addition the change of coordinate
\begin{equation}
u=-\ln w + \sum_i\frac{1}{2}\ln\big(2(z-a_i)\big),
\end{equation}
brings the metric into a familiar form \cite{Hitchin}.
\begin{equation}
\mathrm{d}s^2=\gamma\mathrm{d}z\mathrm{d}\bar{z} + \gamma^{-1} \left( \frac{\mathrm{d}w}{w}+\bar{\delta}\mathrm{d}z \right) \left( \frac{\mathrm{d}w}{w}+ \delta\mathrm{d}\bar{z} \right),
\end{equation}
where
\begin{equation}
\gamma = \sum_i\frac{1}{r_i},
\end{equation}
\begin{equation}
\delta =  \sum_i\frac{x-b_i-r_i}{2(\bar{z}-\bar{a}_i)r_i},
\end{equation}
\begin{equation}
w\bar{w}=\prod_i(x-b_i+r_i).
\end{equation}

Since we would like to calculate the effect of blow-up on the volume of a ball around the origin of $\mathbb{C}^2/\mathbb{Z}_N$, we wish to relate the coordinates $\{ z,\bar{z}, x, y\}$ to the ordinary Cartesian coordinates for flat space.  Let us do this from first principles. Consider the ``blown-down'' version of $U$, which is actually flat space. This is constructed from
\begin{equation}
\mathcal{F}=N\left( r-x\ln (x+r)+\frac{1}{2}x\ln (4z\bar{z}) \right),
\end{equation}
where
\begin{equation} \label{r}
r=\sqrt{x^2+4\lvert z\rvert^2}.
\end{equation}
Using \eqref{Legendre} and \eqref{u} we find the K\"ahler potential is simply
\begin{equation} \label{coord2}
\mathcal{K}=Nr.
\end{equation}
Hence, $Nr$ corresponds to the square of the usual radius in flat space. We can also derive the relation
\begin{equation}
u+\bar{u}=\frac{N}{2}\ln \left( \frac{r-x}{r+x} \right),
\end{equation}
which leads to
\begin{equation}
r=2\lvert z \rvert \cosh \left( \frac{u+\bar{u}}{N} \right),
\end{equation}
\begin{equation}
x=-2\lvert z \rvert \sinh \left( \frac{u+\bar{u}}{N} \right).
\end{equation}
Now flat space Cartesian coordinates $z_1$ and $z_2$ satisfy
\begin{equation} \label{flat}
\mathcal{K}=\lvert z_1 \rvert^2 + \lvert z_2 \rvert^2
\end{equation}
and so identifying \eqref{coord2} and \eqref{flat} we can come up with the following holomorphic transformation relating the two sets of coordinates.
\begin{equation} \label{trans}
z_1=\sqrt{Nz}e^{\frac{u}{N}}, \: \: z_2=\sqrt{Nz}e^{-\frac{u}{N}}.
\end{equation}
The coordinates $z_1$ and $z_2$ are unrestricted, and from this fact and \eqref{trans} we can infer the ranges of the coordinates $\{ x, y, z, \bar{z} \}$. We find that $x$ and $z$ are unrestricted, whilst $y$ is periodic with period $4\pi$.

We would like to compute the volume of a ball around the origin of $U$. First however, we need to define a radial coordinate analogous to $r$ in equation \eqref{r}. The most sensible choice is to define
\begin{equation} \label{r1}
r\equiv \sqrt{(x-\tilde{b})^2+4\lvert z-\tilde{a} \rvert^2},
\end{equation}
where tildes denote mean values over the index $i$. Having done this, we can perform the integration \eqref{integral} over the region $0\leq r\leq R$. Since we will ultimately be interested in the small blow-up limit, let us assume that $R$ is much larger than the $\lvert a_i \rvert$ and $\lvert b_i \rvert$, and derive an answer that is correct to lowest non-trivial order in these blow-up moduli. From \eqref{integral} and \eqref{potential}, the contribution from one term in the sum is given by
\begin{equation}
V=\frac{1}{8}\int\frac{1}{\sqrt{(x-b)^2+4(z-a)(\bar{z}-\bar{a})}}\mathrm{d}x\mathrm{d}y\lvert \mathrm{d}z\mathrm{d}\bar{z}\rvert,
\end{equation}
dropping the subscript $i$ for convenience. We make the change of variables $z=u+iv$, $a=u_0+iv_0$ and then $x^\prime=x$, $u^\prime=2u$, $v^\prime=2v$, $c=2u_0$, $d=2v_0$ and carry out the $y$ integration to obtain
\begin{equation}
V=\frac{\pi}{4}\int\frac{1}{\sqrt{(x^\prime-b)^2+(u^\prime-c)^2+(v^\prime-d)^2}}\mathrm{d}u^\prime\mathrm{d}v^\prime\mathrm{d}x^\prime,
\end{equation}
with the range $(x^\prime-\tilde{b})^2+(u^\prime-\tilde{c})^2+(v^\prime-\tilde{d})^2\leq R^2$. It is now straightforward to obtain
\begin{equation}
V=\frac{\pi}{8}\int\left( \sqrt{R^2+f(\theta , \phi )^2 - \boldsymbol{b}^2}+ f(\theta , \phi ) \right)^2\sin\theta\mathrm{d}\theta\mathrm{d}\phi,
\end{equation}
where $\boldsymbol{b}=(b-\tilde{b},\;c-\tilde{c},\;d-\tilde{d})$ and
\begin{equation}
f(\theta , \phi )=(b-\tilde{b})\sin\theta\cos\phi+(c-\tilde{c})\sin\theta\sin\phi+(d-\tilde{d})\cos\theta.
\end{equation}
Finally, we can do this to lowest non-trivial order in $\boldsymbol{b}$ and substitute back for $a$ to find
\begin{equation} \label{vol1}
V=\frac{\pi^2}{2}\left( R^2 - \frac{1}{3}\left( (b-\tilde{b})^2+4\lvert a-\tilde{a} \rvert^2\right)\right)+\mathcal{O}(\lvert\boldsymbol{b}\rvert^3).
\end{equation}
When we sum \eqref{vol1} over all moduli we obtain the result that, correct to second order in the $a_i$ and $b_i$,
\begin{equation} \label{blowvol}
\mathrm{vol}_U(r=0,R)=\frac{\pi^2}{2}\bigg(NR^2-\frac{N}{3}(\mathrm{var}_i\{b_i\} + 2\,\mathrm{var}_i\{\mathrm{Re}\, a_i\} + 2\,\mathrm{var}_i\{\mathrm{Im}\,a_i\}) \bigg).
\end{equation}
Here var refers to the variance, with the usual definition:
\begin{equation}
\mathrm{var}_i\{X_i\}=\frac{1}{N}\sum_i(X_i-\tilde{X})^2.
\end{equation}
\\

The Gibbons-Hawking space that we have been discussing approaches flat space asymptotically. However, what we really need for our construction of $G_2$ manifolds are smoothed versions of this space which become exactly flat for sufficiently large radius. We now describe how our previous results generalise to a space $U$ which interpolates between Gibbons-Hawking space at small radius and flat space at large radius.

The smoothed version of $\mathcal{F}$ is given by
\begin{equation} \label{smoothF}
\mathcal{F}=\sum_{i=1}^{N}\bigg(r_i-x_i\mathrm{ln}(x_i+r_i)+\frac{x_i}{2}\ln(4z_i\bar{z}_i)\bigg),
\end{equation}
where now
\begin{equation}
x_i=x-\epsilon b_i, \: \: z_i=z-\epsilon a_i,
\end{equation}
\begin{equation}
r_i=\sqrt{x_i^2+4\lvert z_i\rvert^2}.
\end{equation}
Here $\epsilon$ is the smoothing function, dependent on the radius
\begin{equation}
r\equiv \sqrt{(x-\tilde{b})^2+4\lvert z-\tilde{a} \rvert^2},
\end{equation}
and satisfying
\begin{equation}
\epsilon (r) = \left\{ \begin{array}{cc}
1 & \mathrm{if} \; r\leq r_0,  \\
0 & \mathrm{if} \; r\geq r_1. \\
\end{array} \right.
\end{equation}
Further, $r_0 $ and $r_1$ are two characteristic radii satisfying
$\lvert a_i\rvert\ll r_0 < r_1$ and $\lvert b_i\rvert\ll r_0$ for each
$i$ while $U$ describes Gibbons-Hawking space for $r<r_0$ and the flat
space $\mathbb{C}^2/\mathbb{Z}_N$ for $r>r_1$.

Although this space interpolates between two hyperk\"ahler spaces it
is not a hyperk\"ahler space by itself. Accordingly the forms
$\omega^2$ and $\omega^3$ are no longer co-closed in the ``collar''
region $r\in[r_0,r_1]$. However, this space can be thought of as being
close to hyperk\"ahler as long as the blow-up moduli are sufficiently
small compared to one and the function $\epsilon$ is slowly varying
\cite{Lukas}. Analogously, the $G_2$ structure on $T^3\times U$ is not
actually torsion free, but has small torsion under the same
assumptions.

Let us now work out the volume of the region $r\leq\sigma$, for
$\sigma > r_1$. Considering first the contribution up to some radius
$R$, much larger than the $\lvert a_i \rvert$ and $\lvert b_i \rvert$
but smaller than $r_0$ so that $\epsilon$ is identically 1 on the
region of integration, we have exactly the same result as before
\eqref{blowvol}.  The corresponding contribution coming from a shell
$\rho_1<r<\rho_2$ in which $\epsilon=0$ is
\begin{equation}
V=\frac{N\pi^2}{2}\left( \rho_2^2-\rho_1^2 \right).
\end{equation}
Finally we discuss what happens in the ``collar'' region
$r_0<r<r_1$. Here $U$ looks locally like Gibbons-Hawking space, except
that as one moves outward, away from the origin, the modulus
$\boldsymbol{b}$ is decreasing. Therefore local contributions to the
volume from $\boldsymbol{b}$ become smaller as one moves away from the
origin. We make the observation that the contribution to \eqref{vol1}
from $\boldsymbol{b}$ is independent of $R$. Hence, the volume of a
shell with radii much larger than $\lvert \boldsymbol{b} \rvert$ but
smaller than $r_0$ is independent of $\boldsymbol{b}$ to second order.
We deduce that at second order there can be no contribution from
$\boldsymbol{b}$ to the volume of the collar region. Hence
\eqref{vol1} also holds for $R>r_0$. Hence, the result is identical to
the unsmoothed case, namely that
\begin{equation} \label{finalvol}
\mathrm{vol}_{U}(r=0,\sigma) = \frac{\pi^2}{2}\left(N\sigma^2-\frac{N}{3}\left(\mathrm{var}_i\{b_i\} + 2\mathrm{var}_i\{\mathrm{Re}\: a_i\} + 2\mathrm{var}_i\{\mathrm{Im}\:a_i\}\right) \right) + \mathcal{O}(\lvert\boldsymbol{b}_i\rvert^3).
\end{equation}
Note that this expression is independent of the precise form of the smoothing function $\epsilon$.


\begin{thebibliography}{99}
\bibitem{Comp1} G. Papadopoulos and P. K. Townsend, ``Compactification of D=11 supergravity on spaces of exceptional holonomy,'' Phys. Lett. B \textbf{357} (1995) 300 [arXiv:hep-th/9506150].
\bibitem{Comp2} E. Witten, ``Fermion Quantum Numbers in Kaluza-Klein Theory,'' in T. Appelquist, et al.: Modern Kaluza-Klein Theories, 438-511; (in Shelter Island 1983, Proceedings, Quantum Field Theory and the Fundamental Problems in Physics, 227-277).
\bibitem{Atiyah} M. Atiyah and E. Witten, ``M-theory dynamics on a manifold of G(2) holonomy,'' Adv. Theor. Math. Phys. \textbf{6} (2003) 1 [arXiv:hep-th/0107177].
\bibitem{Anomaly} E. Witten, ``Anomaly Cancellation on $G_2$-Manifolds'', arXiv:hep-th/0108165.
\bibitem{beatriz} B. de Carlos and A. Lukas, ``Non-perturbative Vacua for M-theory on G(2) Manifolds,'' arXiv:hep-th/0409255.
\bibitem{House} T. House and A. Lukas, ``$G_2$ Domain Walls in M-theory'', arXiv:hep-th/0409114.
\bibitem{Bilal} A. Bilal and S. Metzger, ``Anomalies in M-theory on singular $G_2$-manifolds'', arXiv:hep-th/0303243.
\bibitem{witten} C. Beasley and E. Witten, ``A note on fluxes and superpotentials in M-theory compactifications on manifolds of G(2) holonomy,'' JHEP \textbf{0207} (2002) 046 [arXiv:hep-th/0203061].
\bibitem{A} M. Cvetic, G. Shiu and A. M. Uranga, ``Chiral four-dimensional N = 1 supersymmetric Type IIA Orientifolds from intersecting D6-branes,'' Nucl. Phys. B \textbf{615} (2001) 3 [arXiv:hep-th/0107166].
\bibitem{B} B.S. Acharya and E. Witten, ``Chiral Fermions from Manifolds of G(2) Holonomy,'' arXiv:hep-th/0109152.
\bibitem{C} Z.W. Chong, M. Cvetic, G.W. Gibbons, H. Lu, C.N. Pope and P. Wagner, ``General Metrics of G(2) Holonomy and Contraction Limits,'' Nucl. Phys. B \textbf{638} (2002) 459 [arXiv:hep-th/0204064].
\bibitem{E} P. Berglund and A. Brandhuber, ``Matter from G(2) Manifolds,'' Nucl. Phys. B \textbf{641} (2002) 351 [arXiv:hep-th/0205184].
\bibitem{F} K. Behrndt, G. Dall'Agata, D. Lust and S. Mahapatra, ``Intersecting 6-branes from new 7-manifolds with G(2) holonomy,'' JHEP \textbf{0208} (2002) 027 [arXiv:hep-th/0207117].
\bibitem{G} J. Gutowski and G. Papadopoulos, ``Brane Solitons for G(2) Structures in Eleven-Dimensional Supergravity re-visited,'' Class. Quant. Grav. \textbf{20} (2003) 247 [arXiv:hep-th/0208051].
\bibitem{H} B.S. Acharya and B. Spence, ``Flux, Supersymmetry and M-theory on 7-manifolds,'' arXiv:hep-th/0007213.
\bibitem{Friedmann} T. Friedmann and E. Witten, ``Unification Scale, Proton Decay and Manifolds of $G_2$ Holonomy'', arXiv:hep-th/0211269.
\bibitem{Gukov} B.S. Acharya and S. Gukov, ``M-theory and Singularities of Exceptional Holonomy Manifolds'', arXiv:hep-th/0409191.
\bibitem{Metzger} S. Metzger, ``M-theory Compactifications, $G_2$ Manifolds and Anomalies'', arXiv:hep-th/0308085.
\bibitem{Acharya} B.S. Acharya, ``M-theory, Joyce Orbifolds and Super Yang-Mills'', arXiv:hep-th/9812205.
\bibitem{Tamar2} T. Friedmann, ``On the Quantum Moduli Space of M-Theory Compactifications'', arXiv:hep-th/0203256.
\bibitem{He} Y. He, ``$G_2$ Quivers'', arXiv:hep-th/0210127.
\bibitem{Ferretti} G. Ferretti, P. Salomonson and D. Tsimpis, ``D-Brane Probes on $G_2$ Orbifolds'', arXiv:hep-th/0111050.
\bibitem{Morris} A. Lukas and S. Morris, ``Rolling $G_2$ Moduli'', arXiv:hep-th/0308195.
\bibitem{Lukas} A. Lukas and S. Morris, ``Moduli K\"ahler Potential for M-theory on a $G_{2}$ Manifold'', arXiv:hep-th/0305078.
\bibitem{Ono} M. Ono, ``Classification of Lattices with $\mathbb{Z}_{m}$ Symmetry'', Commun. Math. Phys. \textbf{126} (1989) 25.
\bibitem{Joyce} D. Joyce, ``Compact Manifolds with Special Holonomy'', Oxford Mathematical Monographs, Oxford University Press, Oxford 2000.
\bibitem{Bailin} D. Bailin and A. Love, ``Orbifold Compactifications of String Theory'', Phys. Rep. \textbf{315} (1999) 285.
\bibitem{d}  L. Dixon, J.A. Harvey, C. Vafa and E. Witten, ``Strings on Orbifolds'', Nucl. Phys. B \textbf{261} (1985) 678.
\bibitem{Dixon} L. Dixon, J.A. Harvey, C. Vafa and E. Witten, ``Strings on Orbifolds (II)'', Nucl. Phys. B \textbf{274} (1986) 285.
\bibitem{Hitch1} N.J. Hitchin, A. Karlhede, U. Lindstrom, and M. Rocek, ``Hyperk\"ahler Metrics and Supersymmetry'', Commun. Math. Phys. \textbf{108} (1987) 535.
\bibitem{Hitchin} N.J. Hitchin, ``Polygons and Gravitons'', Math. Proc. Camb. Phil. Soc. \textbf{85} (1979) 465.
\bibitem{GibbHawk} G.W. Gibbons and S.W. Hawking, ``Gravitational Multi-Instantons'', Phys. Lett. B \textbf{78} (1978) 430.
\bibitem{Strominger} A. Strominger, ``Yukawa Couplings in Superstring Compactification,'' Phys. Rev. Lett. \textbf{55} (1985) 2547.
\bibitem{Candelas} P. Candelas and X. de la Ossa, ``Moduli Space of Calabi-Yau Manifolds,'' Nucl. Phys. B \textbf{355} (1991) 455.
\end{thebibliography}
\end{document}